\documentclass[twocolumn,showpacs,preprintnumbers,amsmath,amssymb]{revtex4}

\usepackage{graphicx}
\usepackage{dcolumn}
\usepackage{bm}
\usepackage{tabularx}
\usepackage{ulem} 
\newcolumntype{M}{>{\centering\arraybackslash}m{1.85cm}}
\usepackage[export]{adjustbox}
\usepackage{float}
\usepackage{color}   
\usepackage{hyperref}
\hypersetup{
    colorlinks=true, 
    linktoc=all,     
    linkcolor=blue,  
}

\makeatletter
\newcommand{\colorcaption}[2][]{%
  \begingroup%
  \renewcommand{\@caption@fignum@sep}{ (Color online). }%
  \caption[#1]{#2}%
  \endgroup%
}
\makeatother
\newcommand\T{\rule{0pt}{3ex}}       
\newcommand\B{\rule[-1.5ex]{0pt}{0pt}} 

\begin{document}

\title{Second-forbidden nonunique $\beta^-$ decays of $^{59,60}$Fe: Possible candidates for $g_{\rm A}$ 
sensitive electron spectral-shape measurements}
\author{ Anil Kumar$^{1}$\footnote{akumar5@ph.iitr.ac.in}, Praveen C. Srivastava$^{1}$\footnote{Corresponding author: praveen.srivastava@ph.iitr.ac.in} and Jouni Suhonen$^{2}$\footnote{jouni.t.suhonen@jyu.fi}}
\address{$^{1}$Department of Physics, Indian Institute of Technology Roorkee, Roorkee 247667, India}
\address{$^{2}$University of Jyvaskyla, Department of Physics, P.O. Box 35 (YFL), FI-40014, 
University of Jyvaskyla, Finland}

\date{\hfill \today}
\begin{abstract}
In  this work, we present a theoretical study of the electron spectral shapes for
the second-forbidden nonunique $\beta^-$-decay transitions 
$^{59}\textrm{Fe}(3/2^-)\to\,^{59}\textrm{Co}(7/2^-)$ and
$^{60}\textrm{Fe}(0^+)\to\,^{60}\textrm{Co}(2^+)$ in the framework of the nuclear shell 
model. We have computed the involved wave functions by 
carrying out a complete $0\hbar\omega$ calculation in the full $fp$ model space
using the KB3G and GXPF1A effective interactions. When compared with the available 
data, these interactions predict the low-energy spectra and electromagnetic 
properties of the involved nuclei quite successfully. This success paves the way
for the computations of the $\beta$-decay properties, and comparison with the available 
data. We have computed the electron spectral shapes of the mentioned decay transitions
as functions of the value of the weak axial coupling $g_{\rm A}$.
By comparing these computed shapes with the measured spectral shapes
allows then to extract the effective value of $g_{\rm A}$ for these decay transitions. This
procedure, coined the spectrum-shape method (SSM) in several earlier studies, complements the 
method of determining the value of $g_{\rm A}$ by reproducing the (partial) half-lives of decay 
transitions. Here we have enhanced the original SSM by  constraining 
the value of the relativistic vector matrix element, 
$^V\mathcal{M}^{(0)}_{KK-11}$, using the conserved vector-current hypothesis (CVC) as a starting
point.
 We hope that this finding would be a strong incentive to measure the spectral shapes in the future.   
\end{abstract}

\maketitle

\section{Introduction}\label{Introduction}

The nuclear $\beta$ decay can be considered as a mutual interaction between the hadronic 
and leptonic current mediated by a massive $W^{\pm}$ vector bosons. These currents can be 
expressed as mixtures of the vector and axial-vector contributions. 
The values of the weak coupling constants enter the theory of $\beta$-decay when the 
hadronic current is renormalized at the nucleon level \cite{Zuber2004}. The free-nucleon 
value of the vector coupling $g_{\rm V}=1.00$ and axial-vector coupling $g_{\rm A}=1.27$ 
derive from the conserved vector-current (CVC) hypothesis and the partially conserved 
axial-vector-current (PCAC) hypothesis, respectively \cite{Commins2007}. The value of 
$g_{\rm A}$ is affected  inside nuclear matter by nuclear many-body, delta-nucleon and mesonic 
correlations. The effect of these corrections on the bare value of $g_{\rm A}$ can be represented 
as an effective value $g_{\rm A}^{\rm eff}=qg_{\rm A}$, where $q$ is a quenching factor.
The effective value of $g_{\rm A}$ plays an important role when data on astrophysical processes, 
single beta decays and double beta decays are to be reproduced by nuclear many-body calculations.
In the single $\beta$ decays, the decay rate depends on the second power of $g_{\rm A}$, while 
to the forth power for $\beta\beta$ decays \cite{Suhonen1998,Maalampi2013}. A comprehensive 
review of the $g_{\rm A}$ problem in $\beta$ and $\beta\beta$ decays is reported in 
Ref. \cite{Suhonen2017}. A recent review of the theoretical and experimental status for 
the single and double $\beta$ decay is given in \cite{Ejiri2019}.

Different methods have been used to extract information on the effective value of $g_{\rm A}$. 
One possibility is the half-life method where the computed and
experimental (partial) $\beta$-decay half-lives are matched by varying the value of $g_{\rm A}$.
This method has been used for the allowed, forbidden, and two-neutrino double $\beta$ decays in 
the framework of the proton-neutron quasiparticle random-phase approximation (pnQRPA)
\cite{Ejiri2014,Ejiri2015, Suhonen2013,Suhonen2014, Faessler2008,Delion2014,Pika2015,Deppisch2016}, 
the nuclear shell model (NSM) 
\cite{Wildenthal1983,Pinedo1996,Caurier2012,Vikas2016,Vikas22016,Archana2018,Anil2020,Joel2020}, 
and the interacting boson model (IBM) \cite{Barea2015,Barea22015,Yoshida2013}.
All these studies have shown that a quenched value of $g_{\rm A}$ is needed in order to reproduced 
the experimental observations.

At this point it is worth pointing out that in the present and the above-referenced
works an important component of nuclear structure has been omitted. This component are
the two-body (meson-exchange) currents which go beyond the usually adopted impulse
approximation of the weak nuclear decays. These currents have long been known to be
active in the first-forbidden non-unique pseudoscalar $0^-\leftrightarrow 0^+$ beta-decay 
transitions governed by the relativistic axial rank-0 matrix element. For these transitions
an enhancement, instead of quenching, has been recorded (see, e.g., 
\cite{Suhonen2017,Warburton1991,Kostensalo2018,Suhonen2019}). The two-body currents are
also active in the allowed Gamow-Teller transitions, and taking them into account reduces
drastically the need for the $g_{\rm A}$ quenching, as demonstrated for the very
light nuclei in \cite{Pastore2018,King2020} and light nuclei and 
light medium-mass nuclei in \cite{Gysbers2019}. The studies reproduce well beta-decay
rates using an unquenched value of $g_{\rm A}$ by means of including additional
nuclear correlations and the two-body currents, as discussed in the 
review \cite{Engel2017}.

Regarding the role of many-body correlations and the choice of model space, the treatment
of Gamow-Teller transitions and forbidden transitions may need a different degree of 
sophistication. In particular, for the Gamow-Teller transitions a $0\hbar\omega$ calculation,
within one single oscillator major shell, is a good approximation, but for the
second-forbidden beta transitions, discussed in the present work, the situation is not 
that clear. For the second-forbidden transitions the transition operators tend to
pick $2\hbar\omega$ contributions neglected in the present calculations. As discussed
below, this is reflected in the value of a key relativistic vector matrix element, which
in our calculations vanishes, but should have a non-zero value according to the 
CVC hypothesis. The evidence from the electric quadrupole ($E2$) transitions is that
the $2\hbar\omega$ contributions can be accounted for by the use of effective charges
that in the standard form enhance the isoscalar contribution without modifying the
isovector one, which is relevant for beta decay. However, there are no estimates about the
role of the $2\hbar\omega$ correlations in the context of spin-quadrupole transitions,
mediated by the axial second-forbidden operators relevant for the present calculations.

In Ref. \cite{mika2016}, for the first time, another method to determine the effective value 
of $g_{\rm A}$, coined the spectrum-shape method (SSM), was introduced. In the SSM, an electron
spectral shape is computed as a function of the value of $g_{\rm A}$, and then compared with 
the corresponding experimental shape, in order to find the effective $g_{\rm A}$ for which 
the computed spectral shape matches the experimental one. This method is applicable to 
forbidden nonunique $\beta$ decays since the associated electron spectra depend on the 
details of nuclear structure. In Ref \cite{mika2016} also the next-to-leading-order (NLO) 
corrections were included in the spectral shape. 

In Ref. \cite{mika2016} the shape of the electron spectrum for the forth-forbidden nonunique 
$\beta^-$ decay of $^{113}$Cd was computed under the framework of the microscopic 
quasiparticle-phonon model (MQPM) and the NSM. This work was extended in Ref. \cite{mika2017} 
to include a comparison with the results of  the third nuclear model, IBM.
In \cite{mika2017} the computed spectral shapes were compared with the measured one of 
Belli et al. \cite{Belli2007}, and the closest match was found
for the ratio of $g_{\rm A}/g_{\rm V}\approx0.92$ for all three nuclear models.

In continuation, electron spectral shapes for several experimentally interesting odd-$A$ 
nuclei (MQPM and NSM calculations) and even-$A$ nuclei (NSM calculations) were studied for their
$g_{\rm A}$ dependence in Refs. \cite{joel12017,joel22017}.
In Refs. \cite{mika2016,joel12017,joel22017} it was found that the spectral shapes for the 
$\beta$ decays of $^{87}$Rb, $^{94}$Nb, $^{98}$Tc, $^{99}$Tc, $^{113}$Cd, and $^{115}$In depend 
strongly on the effective value of $g_{\rm A}$, thus rendering these decays as excellent 
candidates for applications of the SSM. 

Recently, in Ref. \cite{Anil2020PRC} the  electron spectral shapes of the second-forbidden 
nonunique $\beta^-$ decays of $^{24}$Na and $^{36}$Cl were studied. It was found that the 
potential of the SSM could be enhanced by constraining the magnitude of the small relativistic
vector matrix element, $^V\mathcal{M}_{211}^{(0)}$, by the measured decay 
half-life. This then leads to a consistent treatment of both the half-life and the spectral 
shape. Another approach, using the CVC to constrain the value of $^V\mathcal{M}_{211}^{(0)}$, 
was used for the second-forbidden nonunique $\beta^-$ decay of 
$^{20}$F in \cite{Kirsebom2019PRC,Kirsebom2019PRL}.

In order to extend the SSM studies, we have tried to find new candidates for the application 
of the SSM by scanning through the $fp$ shell and doing spectral-shape calculations using 
well established shell-model Hamiltonians. In this region of the nuclear chart
we have found two possible candidate transitions, 
$^{59}\textrm{Fe}(3/2^-)\to\,^{59}\textrm{Co}(7/2^-)$ and
$^{60}\textrm{Fe}(0^+)\to\,^{60}\textrm{Co}(2^+)$,
which correspond to second-forbidden nonunique $\beta^-$ decays with a $Q$ values of 1.565 MeV
and 0.178 MeV, branching ratios of 0.18(4)\% and 100\%, and partial half-lives of 67.72 years
and $2.62\times 10^6$ years, respectively. Here we study these
transitions by carrying out complete $0\hbar\omega$ shell-model calculations in a full $fp$ 
model space by using the well-established two-body interactions 
GXPF1A \cite{Honma2004,Honma2005} and KB3G \cite{Poves2001}. Here it is worth mentioning 
that results for the $^{59}$Fe electron spectra, corresponding to the GXPF1A interaction, have
previously been reported in Ref. \cite{Suhonen2019}. However, there the calculations
were done in a very restricted model space and the drawn conclusion about the sensitivity to the
SSM analysis differs from the conclusions of the present calculations. 

In the present work, we enhance the original SSM in two different ways: either by i) constraining 
the value of the relativistic vector matrix element, $^V\mathcal{M}^{(0)}_{KK-11}$, using 
the CVC as described in \cite{behrens1982}, or ii) 
using the value of the matrix element as a fit parameter to reproduce the partial half-life
of the decay transition. The former method requires an ``ideal''
nuclear-structure calculation, but in the present calculations we have to confine ourselves 
to the impulse approximation and a finite single-particle space, making the method i) only 
approximative in the present case. Nevertheless, the results of method i) can serve as rough 
estimates of the order of magnitude and sign of $^V\mathcal{M}^{(0)}_{KK-11}$, thus giving
the method ii) a good starting point for the half-life fit. By using these two methods, 
we want to test how sensitively the prediction for the spectral 
shape depends on the method used to obtain it.

In order to test the predictive power of our computed  nuclear wave functions, we have calculated 
the energy spectra for low-lying states as well as other spectroscopic properties of both the 
parent and daughter nuclei involved in the studied $\beta$-decay transitions. All these computed 
quantities turn out to be in good agreement with the available data. After all these comparisons 
we finally compute the electron spectral shapes as functions of the value of $g_{\rm A}$.
It turns out that the spectral shapes obtained by using the method ii) dependent rather 
sensitively on the value of $g_{\rm A}$. Since in the case of the present, non-ideal, calculations 
the method ii) is to be viewed as more reliable, the enhanced SSM can be considered as
a powerful probe for the studied transitions. We hope that these findings serve as a strong 
incentive to measure the involved spectral shapes in the future.

The current article is organized as follows. In Sec. \ref{Formalism} we give a short overview 
of the adopted theoretical formalism for $\beta$-decay. Results and discussions are reported in 
Sec. \ref{Results} and, finally, in Sec. \ref{Conclusions}, we conclude.

\section{Theoretical Formalism} \label{Formalism}
 In Sec. \ref{beta},  we  discuss  the theory of forbidden nonunique $\beta^-$ decays, and
 the shape of the electron spectra.  In Sec. \ref{models} we give the details about
 the valence space and effective Hamiltonian used in the present work.
 
 \subsection{$\beta$-decay theory} \label{beta}

In the literature, the theoretical framework of the $\beta$ decay is well established in the 
book by Behrens and B$\ddot{\text{u}}$hring \cite{behrens1982} (see also Ref. \cite{hfs1966}). 
We have used the streamlined version of the formalism for the forbidden nonunique 
$\beta^-$-decay theory from Refs. \cite{mst2006,mika2017}. 
To simplify the $\beta^-$-decay theory, we have used the impulse  approximation, in which at 
the exact moment of the decay, only the decaying nucleon feels the weak interaction and 
the strong interactions with the remaining $A-1$ nucleons are ignored \cite{suhonen2007}.
For the $\beta^-$-decay process, described as an effective point-like interaction vertex 
with an effective coupling constant $G_{\text{F}}$, called Fermi coupling constant, the 
probability of the emitted electron to have a kinetic energy between $W_e$ and $W_e+dW_e$ is 
given by  

\begin{equation} \label{eq1}
\begin{split}
P(W_e)dW_e & = \frac{G_\text{F}^2}{(\hbar{c})^6}\frac{1}{2\pi^3\hbar}C(W_e)p_ecW_e(W_0-W_e)^2 \\
 & \times{F_0(Z,W_e)dW_e},
\end{split}
\end{equation}
where $W_0$ is the endpoint energy of the $\beta$ spectrum,   the factor $F_0(Z, W_e)$ is 
the Fermi function, and  $Z$ is the proton number of the daughter nucleus. The $p_e$ 
and $W_e$ are the momentum and energy of the emitted electron, respectively.  Furthermore, 
the shape factor $C(W_e)$ contains the nuclear-structure information.  

The partial half-life of the decay process can be written as

\begin{eqnarray}\label{hf1}
t_{1/2}=\frac{\text{ln}(2)}{\int_{m_ec^2}^{W_0}{P(W_e)dW_e}}=\frac{\kappa}{\tilde{C}},
\end{eqnarray}
where $m_e$ is the rest mass of the electron, $\tilde{C}$ is the dimensionless integrated shape 
function, and the updated \cite{Patrignani} value  of  constant $\kappa$ is

\begin{eqnarray}
\kappa=\frac{2\pi^3\hbar^7\text{ln(2)}}{m_e^5c^4(G_\text{F}\text{Cos}\theta_\text{C})^2}=
6289~\mathrm{s},
\end{eqnarray}
where the $\theta_\text{C}$ is the Cabibbo angle. To simplify the formalism it is usual to 
adopt the unitless scaled kinematics quantities $w_0=W_0/m_ec^2$, $w_e=W_e/m_ec^2$, 
and $p=p_ec/m_ec^2=\sqrt{(w_e^2-1)}$.  
Using the unitless quantities the dimensionless integrated shape function reads 

\begin{eqnarray} \label{tc}
\tilde{C}=\int_1^{w_0}C(w_e)pw_e(w_0-w_e)^2F_0(Z,w_e)dw_e.
\end{eqnarray}

 The general form of the shape factor $C(w_e)$ can be expressed as

\begin{eqnarray} \label{eq2}
C(w_e)  = \sum_{k_e,k_\nu,K}\lambda_{k_e} \Big[M_K(k_e,k_\nu)^2+m_K(k_e,k_\nu)^2 \nonumber\\
    -\frac{2\gamma_{k_e}}{k_ew_e}M_K(k_e,k_\nu)m_K(k_e,k_\nu)\Big],
\end{eqnarray}
where the $k_e$ and $k_\nu$ (both are running through 1, 2, 3,...) are the positive integers 
emerging from the partial-wave expansion of the leptonic wave functions, and $K$ is the 
order of the forbiddenness. The quantities $M_K(k_e,k_\nu)$ and $m_K(k_e,k_\nu)$  contain  
all the nuclear-structure information in the form of 
different nuclear matrix elements (NMEs) and leptonic phase-space factors. More 
information on these expressions can be found in Ref. \cite{behrens1982} (also given in 
Ref. \cite{mika2017}). Here $\gamma_{k_e}=\sqrt{k_e^2-(\alpha{Z})^2}$ and the quantity 
$\lambda_{k_e}={F_{k_e-1}(Z,w_e)}/{F_0(Z,w_e)}$ is the  Coulomb function, where 
${F_{k_e-1}(Z,w_e)}$ is the generalized Fermi function \cite{mst2006,mika2017}.

The NMEs contains all the nuclear-structure information in the form

\begin{align}
\begin{split}
^{V/A}\mathcal{M}_{KLS}^{(N)}(pn)(k_e,m,n,\rho)& \\=\frac{\sqrt{4\pi}}{\widehat{J}_i}
\sum_{pn} \, ^{V/A}m_{KLS}^{(N)}(pn)(&k_e,m,n,\rho)(\Psi_f|| [c_p^{\dagger}
\tilde{c}_n]_K || \Psi_i),
\label{eq:ME}
\end{split}
\end{align}
where the quantities ${^{V/A}m_{KLS}^{(N)}}(pn)(k_e,m,n,\rho)$ are called the single-particle 
matrix elements (SPMEs), which characterize the properties of the transition operators, 
and they are the same for all the nuclear models. In our calculations, the SPMEs are 
computed in basis of harmonic-oscillator wave functions \cite{mika2017,mst2006}. The 
quantities  $(\Psi_f|| [c_p^{\dagger}\tilde{c}_n]_K || \Psi_i)$ are the one-body transition 
densities (OBTDs) between the initial $(\Psi_i)$ and final $(\Psi_f)$ states. The OBTDs contain
the nuclear-structure information and they must be evaluated separately for each nuclear 
model. The summation runs over the proton $(p)$ and neutron $(n)$ single-particle states 
and the ``hat-notation'' reads ${\widehat{J}_i}=\sqrt{2J_i+1}$.
The Coulomb-corrected NMEs are indicated by the notation $(k_e,m,n,\rho)$ when such matrix 
elements exist, and they are characterized by using the integers $m$, $n$, and $\rho$, 
where the order $m$ is the total power of the factors ($m_eR$), ($W_eR$) and ($\alpha{Z}$), 
the number $n$ is the total power of the factors ($W_eR$) and ($\alpha{Z}$) and the number 
$\rho$ the power of the factor ($\alpha{Z}$). Here $R$ is the nuclear radius, and  $L$ and 
$S$ are the total orbital angular momentum and spin of the leptons, respectively.  The label $N$ refers to the power of the $qR$ when expanding the form factors, where $q$ is the momentum transfer. 

The shape factor depends on the weak coupling constants $g_{\rm V}$  and $g_{\rm A}$. So the 
shape factor can be decomposed into vector, axial-vector and mixed vector-axial-vector 
components \cite{mika2017,mika2016,joel12017,joel22017} based on the weak coupling constants 
they contain. In this spirit we can write  

\begin{eqnarray}\label{dcmp}
C(w_e)=g_{\rm V}^2C_{\rm V}(w_e)+g_{\rm A}^2C_{\rm A}(w_e)+g_{\rm V}g_{\rm A}C_{\rm VA}(w_e).
\end{eqnarray}

 After integrating Eq. (\ref{dcmp}) with respect to electron kinetic energy, we obtain a 
decomposition of the  dimensionless integrated shape function  (\ref{tc}) in the form.

\begin{eqnarray}\label{intc}
\tilde{C}=g_{\rm V}^2\tilde{C}_{\rm V}+g_{\rm A}^2\tilde{C}_{\rm A}+g_{\rm V}g_{\rm A}\tilde{C}_{\rm VA}.
\end{eqnarray}
The shape factors $C_i$ in Eq. (\ref{dcmp}) depend on the electron kinetic energy,
while after the integration the (partial) shape functions  $\tilde{C}_i$ in Eq. (\ref{intc}) 
are just constant numbers.

\subsection{Adopted model space and  Hamiltonians}\label{models}
For the calculations of the OBTDs, needed for the evaluation of the NMEs contained in the 
$\beta$-decay amplitudes, we need to choose a nuclear model.  In the present  work, the wave 
functions of the initial and final states were computed by using the nuclear shell model (NSM).   
The shell-model wave functions and OBTDs were computed using the nuclear shell-model code 
NuShellX@MSU \cite{brown2014} with the well-known effective interactions KB3G and GXPF1A in 
the full $fp$ model space. In the present studies, we have parformed complete 
$0\hbar\omega$ calculations in the full $fp$ model space.

\section{Results and discussions}\label{Results}
In this section  we  present our  computed results of low-lying energy spectra, spectroscopic 
properties, shape factors and electron spectra for the second-forbidden nonunique 
$\beta^-$-decay transitions  $^{59}$Fe$(3/2^-)\to\,^{59}$Co($7/2^-$) and 
$^{60}$Fe$(0^+)\to\,^{60}$Co($2^+$).
 
The NSM-computed electron spectrum for the decay of $^{60}$Fe is already available in 
Ref. \cite{joel22017}, but in a heavily truncated model space.
In the present work, we have performed complete $0\hbar\omega$ shell-model calculations 
in a full $fp$ model space with recent well-established  interactions. 
As in the works \cite{mika2016,mika2017,joel12017,joel22017,Suhonen2019}, we have included 
here the next-to-leading-order (NLO) corrections to the shape factor.
In this way, the number of NMEs increases drastically, and in the case of the 
second-forbidden nonunique $\beta^-$ decay, the number of NMEs is increased from 8 to 27 
(see the full details about the NLO corrections in Refs. \cite{mika2016,mika2017}).

\begin{figure*}[bt]
\centering
\includegraphics[width=\columnwidth]{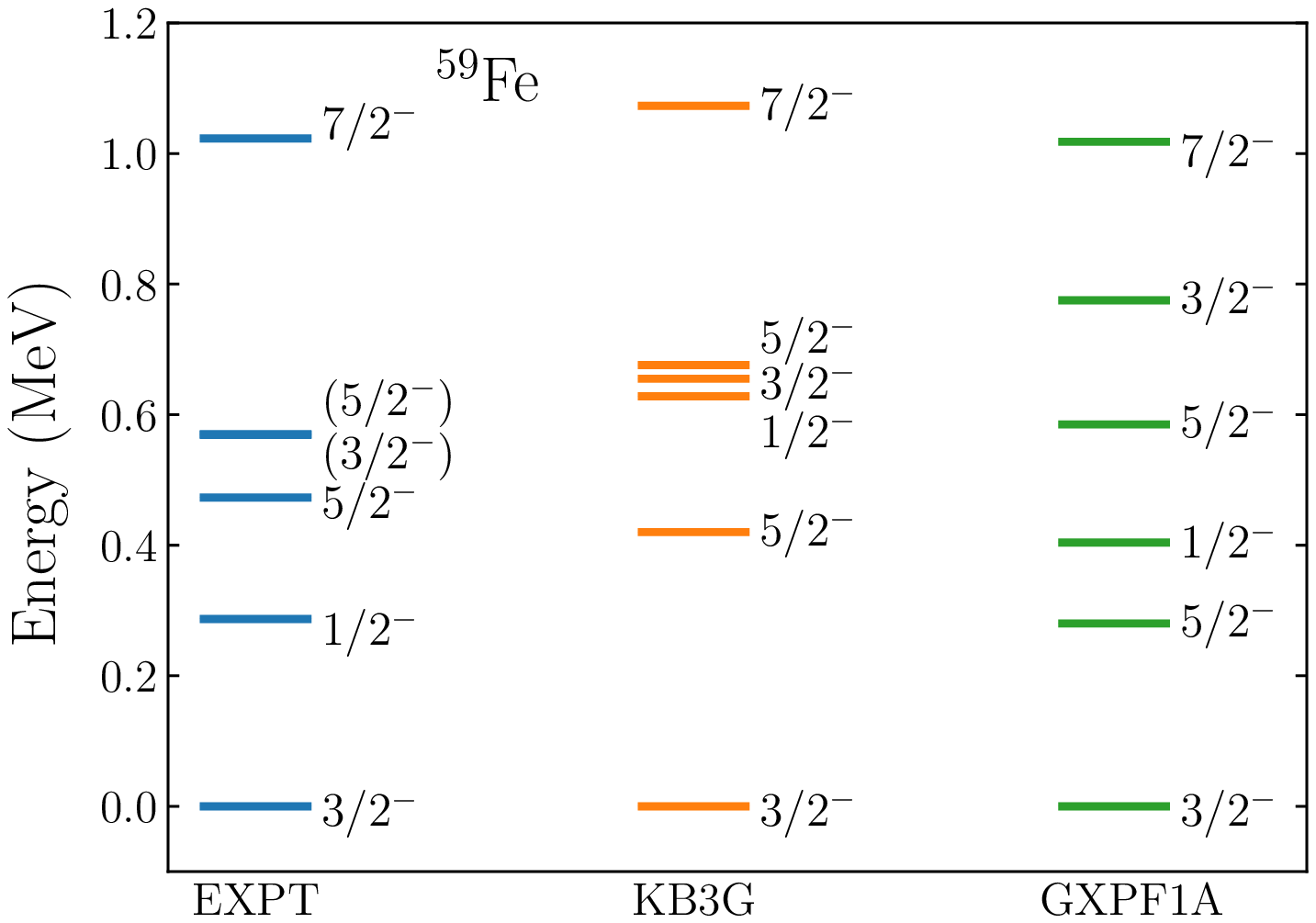}
\includegraphics[width=\columnwidth]{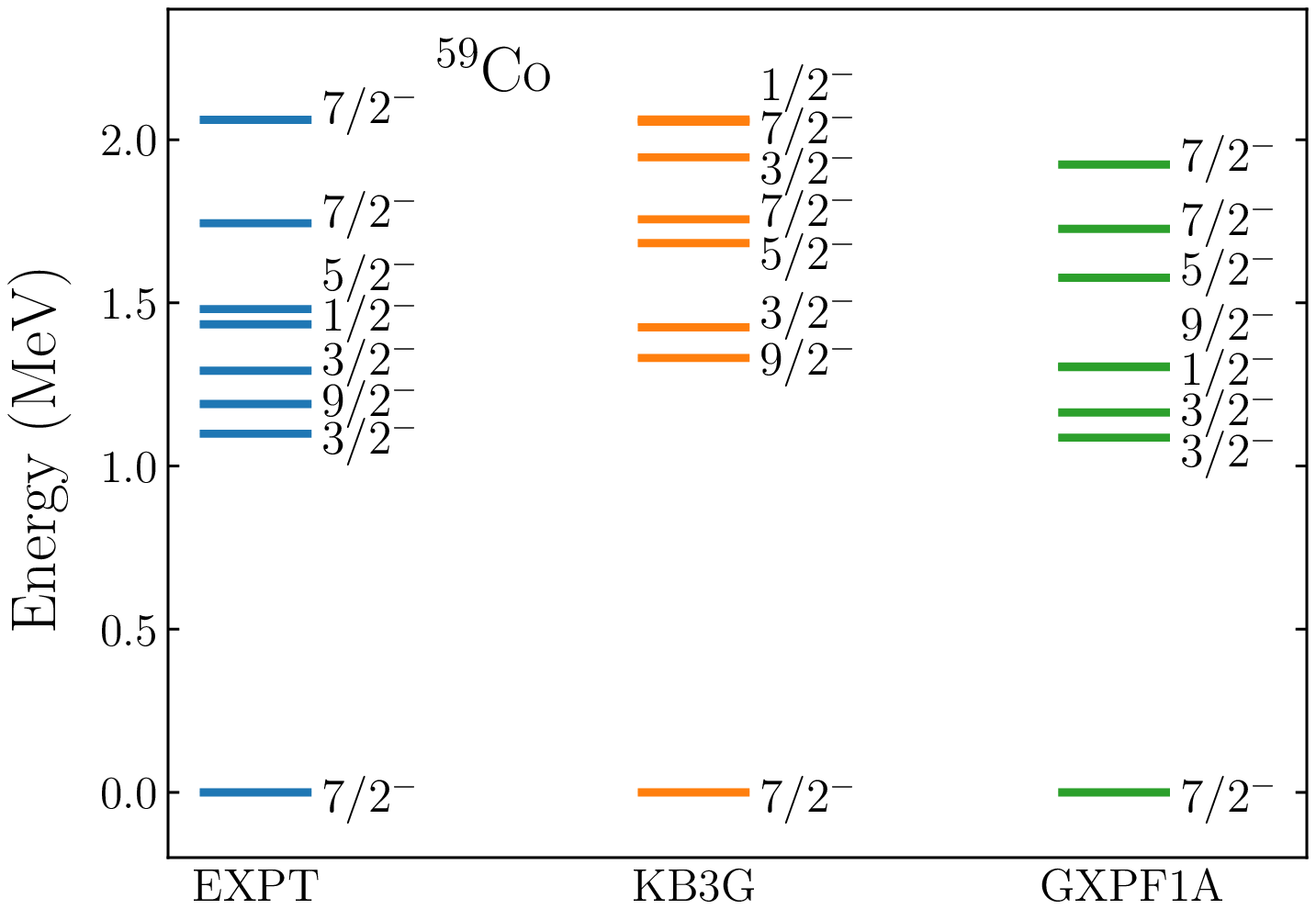}
\caption{Comparison of the KB3G- and GXPF1A-computed energy spectra with the 
experimental \cite{nndc} one for the low-lying energy spectra of $^{59}$Fe and $^{59}$Co.}
\label{fig:Energy_59FeCo}
\end{figure*}
\begin{figure*}[ht]
\centering
\includegraphics[width=\columnwidth]{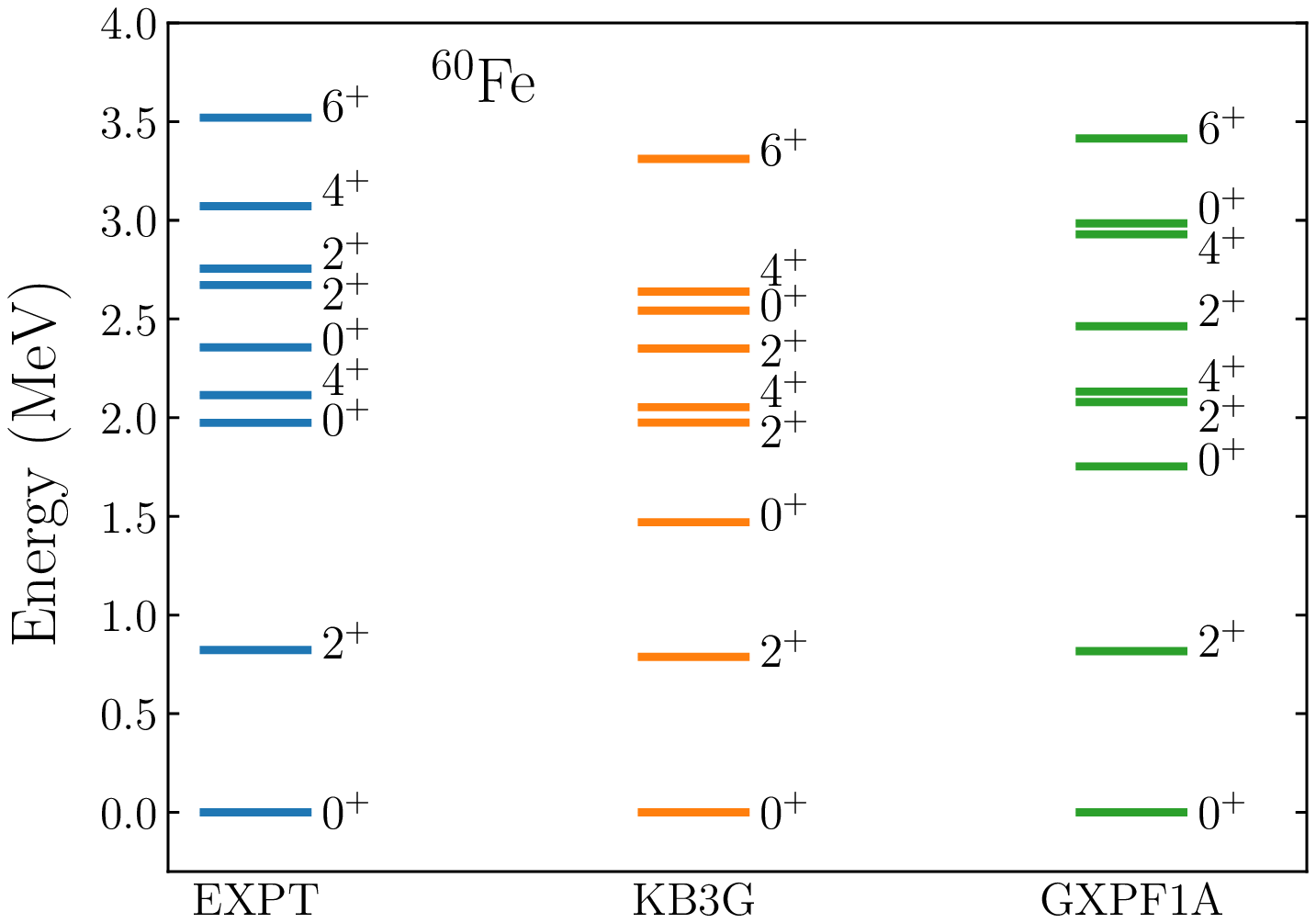}
\includegraphics[width=\columnwidth]{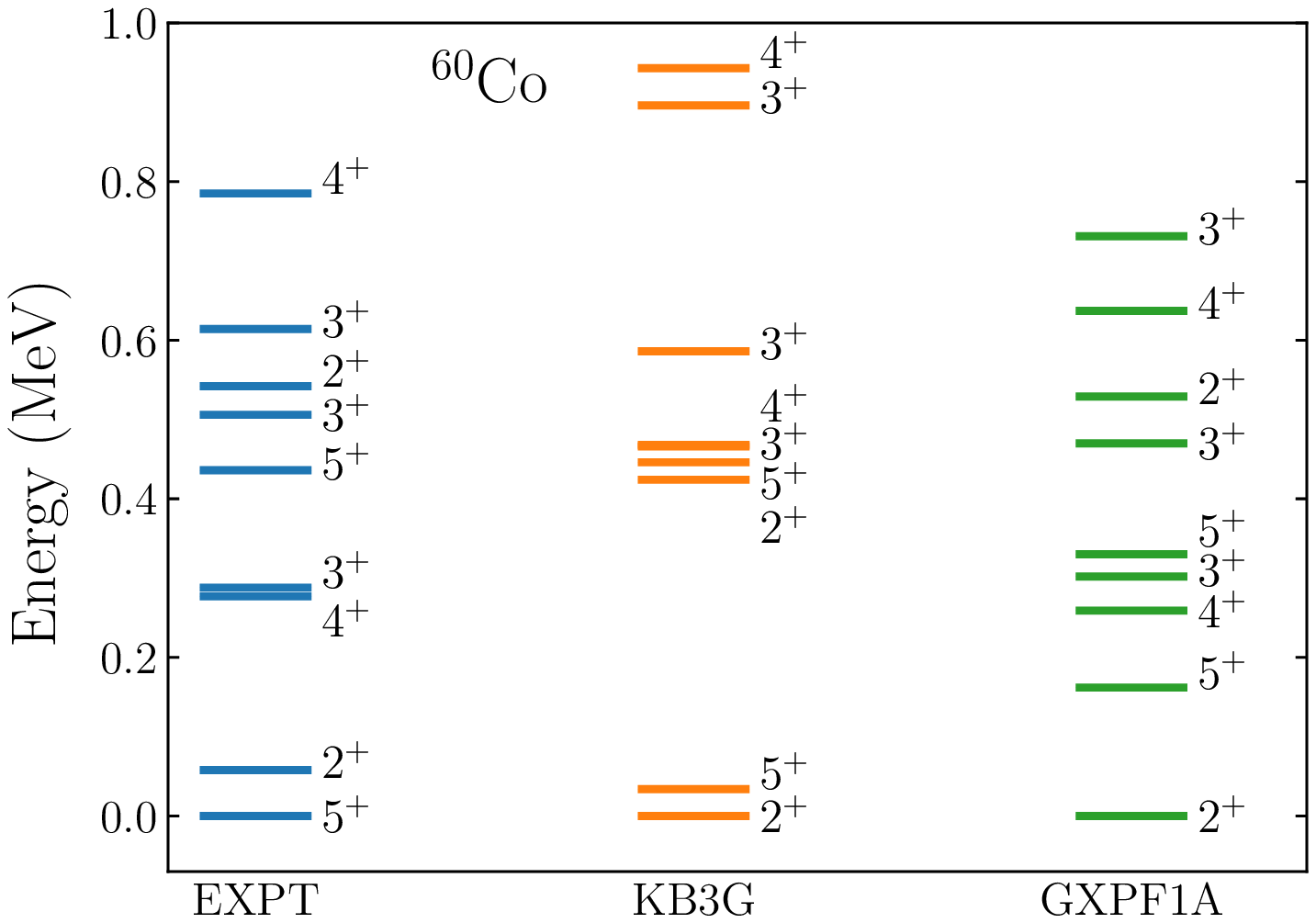}
\caption{Comparison of the KB3G- and GXPF1A-computed energy spectra with the 
experimental \cite{nndc} one for the low-lying energy spectra of $^{60}$Fe and $^{60}$Co.}
\label{fig:Energy_60FeCo}
\end{figure*}

Below we  present low-lying energy spectra (Figs. \ref{fig:Energy_59FeCo} and
\ref{fig:Energy_60FeCo}), spectroscopic properties (Table \ref{table:Qm_moments}
and \ref{table:BE2}), the computed NMEs (Tables \ref{table:mgt}-\ref{table:nmes_CVC}),
electron spectral shapes as functions of the electron kinetic energy
(Figs. \ref{fig:59Fe_spectra_KB3G}-\ref{fig:60Fe_spectra_GXPF1A}), the NLO corrections in the half-life  and electron spectra (Figs.  \ref{fig:60Fe_HL_Corr} and \ref{fig:60Fe_spectra_LO_NLO}), and  decompositions of the integrated shape functions (Table \ref{table:Fe_int_shape}).

\subsection{Low-lying energy spectra and spectroscopic properties}

\begin{table*}[!ht]
\caption{Comparison of the computed quadrupole and magnetic moments with available 
experimental data. For the calculations, we have used  the effective charges $e_p=1.5e$ and 
$e_n=0.5e$, and bare $g$ factors ($g^{\text{eff}}=g^{\text{free}}$). The experimental values 
are taken from \cite{nndc}.}\label{table:Qm_moments}
\begin{ruledtabular}
\begin{tabular}{lccccccc}

  & & &\multicolumn{1}{c} {$Q(eb)$} & \multicolumn{4}{c} {\hspace{27mm} $\mu${($\mu_N$)}}\T\B\\
 \cline{3-5}
 \cline{6-8}
       &  & Expt. &  KB3G  & GXPF1A  & Expt. & KB3G  & GXPF1A \T\B\\
\hline

$^{59}$Fe & $3/2^-$ & N/A      &  +0.21   &  +0.24 & -0.3358 (4) & -0.24  & -0.08 \\

$^{59}$Co & $7/2^-$ & +0.42 (3)  & +0.38  & +0.44 & +4.627 (9)& +4.51 & +4.59  \\

 $^{60}$Fe & $2^+$ & N/A  &  -0.28 & -0.30 & N/A & +1.06 & +1.11 \\

$^{60}$Co & $5^+$ & +0.44 (5)  & +0.44  & +0.51 & +3.799 (8)& +3.66 & +3.95  \\
           & $2^+$ & +0.3 (4)  & +0.23  & +0.26 & +4.40 (9)& +4.60 & +4.26 \\
\end{tabular}
\end{ruledtabular}
\end{table*}

\begin{table}[!ht]
\caption{Comparison of the computed and experimental $B(E2)$ values in W.u. The effective 
charges $e_p=1.5e$ and $e_n=0.5e$ were used. The experimental values are taken 
from \cite{nndc}.}\label{table:BE2}
\begin{ruledtabular}
\begin{tabular}{lcccc}

       & Transitions & Expt. &  KB3G  & GXPF1A \T\B\\
\hline

$^{59}$Fe & $B(E2; 7/2^-\to\, {3/2^-})$ & N/A & 12.36 & 14.42\T\\

$^{59}$Co & $B(E2; 3/2^-\to\, {7/2^-})$ & 9.2 (20) & 6.45 & 0.47\B\\

 $^{60}$Fe & $B(E2; 2^+\to\, {0^+})$ & 13.6 (14) & 15.53 & 18.89 \\

 $^{60}$Co & $B(E2; 4^+\to\, {2^+})$ & N/A & 0.69 &0.57 \\

\end{tabular}
\end{ruledtabular}
\end{table}

We have performed shell-model calculations for the ground state and a few low-lying 
excited states of the $\beta^-$-decay parent and daughter nuclei of the studied transitions 
by using the KB3G and GXPF1A interactions in a full $fp$ model space, performing complete 
$0\hbar\omega$ shell-model calculations. 

The computed energy spectra of low-lying states in $^{59}$Fe and $^{59}$Co are presented 
in Fig. \ref{fig:Energy_59FeCo} and  compared with the available experimental data. As 
seen in Fig. \ref{fig:Energy_59FeCo}, the computed ground states are correctly reproduced 
by the KB3G and GXPF1A interactions for $^{59}$Fe and $^{59}$Co.
Also the computed excited states are in the right energy regions, though some inversions 
in the relative ordering of the states occur for both interactions.
In Fig. \ref{fig:Energy_60FeCo}, we show the low-energy spectra of $^{60}$Fe and $^{60}$Co 
and their comparison with the experimental data.
 As seen in Fig. \ref{fig:Energy_60FeCo}, the computed first $2^+$ state for $^{60}$Fe is 
obtained at 0.788 and 0.817 MeV corresponding to the KB3G and GXPF1A interactions, 
respectively, while the experimental value is 0.823 MeV.
For $^{60}$Co, the KB3G and GXPF1A interactions give a $2^+$ state as the ground state while 
the experimental ground state is $5^+$. From the KB3G calculation, the first excited $5^+$ 
state is obtained at 0.034 MeV while the GXPF1A calculation places it at 0.162 MeV.  
For both $^{60}$Fe and $^{60}$Co the computed spectra contain the spin-parities of the 
experimental spectra in roughly the right energy ranges, but sometimes in inverted orderings.

The computed results for quadrupole and magnetic moments are shown in 
Table \ref{table:Qm_moments} and compared with available experimental data. 
In most cases, both moments are well reproduced by both interactions.
The computed $B(E2)$ values are presented in Table \ref{table:BE2}. Our computed results 
of electromagnetic properties are in good agreement with the available experimental data.  
Present calculations also reproduce correctly the signs of the quadrupole and magnetic moments. 
Overall, the spectroscopic properties of the presently
discussed $fp$-shell nuclei are fairly well reproduced by both the adopted effective
interactions. This gives us confidence for a successful computation of the $\beta$-decay
properties, discussed in the following section.

\begin{table*}[!ht]
\caption{Calculated absolute values of the Gamow-Teller ($|\mathcal{M}_{\rm GT}|$) and Fermi 
($|\mathcal{M}_{\rm F}|$) matrix elements of the allowed $\beta^-$ decays from the ground-state 
($3/2^-$) of $^{59}$Fe to the first two excited $3/2^-$ states in $^{59}$Co. The experimental
Gamow-Teller matrix elements are obtained from the measured $\log ft$ values \cite{nndc} by 
assuming the value $g_{\rm A}=1.27$ for the axial coupling strength.}
\label{table:mgt}
\begin{ruledtabular}
\begin{tabular}{lccccc}

   &\multicolumn{3}{c} {$|\mathcal{M}_{\rm GT}|$}  &\multicolumn{2}{c}{$|\mathcal{M}_{\rm F}|$}\T\B\\
 \cline{2-4}
 \cline{5-6}
        &   Expt. &  KB3G  & GXPF1A & KB3G  & GXPF1A \T\B\\
\hline
$3/2^-\to\,3/2^-_1$   &0.0904  & 0.0696   & 0.1586  &7.24$\times{10^{-6}}$& 
1.0$\times{10^{-5}}$ \T\\
$3/2^-\to\,3/2^-_2$   &0.2064 & 0.0677 &  0.0345  & 3.49$\times{10^{-6}}$ & 
2.2$\times{10^{-5}}$ \T\B\\
\end{tabular}
\end{ruledtabular}
\end{table*}

\begin{table*} [!ht]
\leavevmode
\centering
\caption{\label{table:nmes} Leading-order (LO) and next-to-leading-order (NLO) nuclear matrix 
elements (NMEs) of the second-forbidden nonunique $\beta^- $  decays of $^{59,60}$Fe, computed 
using  the KB3G and GXPF1A interactions. The CVC-constrained s-NMEs, 
$^V\mathcal{M}^{(0)}_{KK-11}$(CVC) (second line), are compared with the
half-life fixed s-NMEs (for $g_{\rm A}=0.80$), $^V\mathcal{M}^{(0)}_{KK-11}$(Half-life) (third line).
The Coulomb-corrected NMEs are indicated by ($k_e,m,n,\rho$), when such elements exist. 
The blank spaces denote vanishing NMEs.  } 
\begin{ruledtabular}
\begin{tabular}{lcccc}
Transition &\multicolumn{2}{c} {$^{59}$Fe$(3/2^-)\to\,^{59}$Co($7/2^-)$} 
      &\multicolumn{2}{c}{$^{60}$Fe$(0^+)\to\,^{60}$Co($2^+)$}   \T\B\\
 
 \cline{2-3}
 \cline{4-5}
 
 NMEs & KB3G & GXPF1A  & KB3G & GXPF1A  \T\B \\
\hline 
{\bf LO}\vspace{0.2cm}\T\\
$^V\mathcal{M}^{(0)}_{KK-11}$  &        0     & 0 &0&0\\
$^V\mathcal{M}^{(0)}_{KK-11}$(CVC)  &        0.2164     & 0.1921&  0.2862 &  0.3055 \\
$^V\mathcal{M}^{(0)}_{KK-11}$(Half-life) & 0.1037 & 0.0950 & 0.1143&0.0919   \T\\
$^V\mathcal{M}^{(0)}_{KK0}$    & 13.2021 &11.7172 & 20.3265 & 21.6968\\
\hspace{1cm}$(1,1,1,1)$        & 15.3084 &13.6063&23.6439 & 25.2589 \\
\hspace{1cm}$(2,1,1,1)$        & 14.4779 &12.8727 &22.3782 & 23.9122\\ 

$^A\mathcal{M}^{(0)}_{KK1}$    & 7.7745 & 7.7344&13.3677 & 19.7997 \\
\hspace{1cm}$(1,1,1,1)$        & 8.8039 & 8.7879&15.2409 & 22.8508 \\
\hspace{1cm}$(2,1,1,1)$        & 8.2773 & 8.2694&14.3539 & 21.5871 \\

$^A\mathcal{M}^{(0)}_{K+1K1}$  &   5.6712 & 4.5057 & $--$& $--$\vspace{0.2cm} \vspace{0.2cm} \\
{\bf NLO}\vspace{0.2cm}\\

$^V\mathcal{M}^{(1)}_{KK-11}$  &-0.1383 &-0.1261 &-0.1965 & -0.2135\\
\hspace{1cm}$(1,1,1,1)$        & -0.1321 &-0.1197&-0.1856 & -0.1968\\
\hspace{1cm}$(2,1,1,1)$        & -0.1187 &-0.1074&-0.1664 & -0.1752 \\
\hspace{1cm}$(1,2,1,1)$        & 0.0242 & 0.0223 &0.0351&0.0395\\
\hspace{1cm}$(2,2,1,1)$        & 0.0169 & 0.0156 &0.0245 &0.0276\\
\hspace{1cm}$(1,2,2,1)$        & -0.1563 &-0.1420&-0.2207 &-0.2363 \\
\hspace{1cm}$(2,2,2,1)$        &-0.1527 &-0.1387  &-0.2154 &-0.2303 \\
\hspace{1cm}$(1,2,2,2)$        &-0.1782 &-0.1613 &-0.2498 &-0.2635 \\
\hspace{1cm}$(2,2,2,2)$        &-0.1711 &-0.1547&-0.2396 &-0.2520 \\

$^A\mathcal{M}^{(1)}_{KK1}$    & 9.7384 & 9.5861 &16.3829 &23.3149\\

$^V\mathcal{M}^{(0)}_{KK+11}$  &  9.6599 & 8.8069 &13.8826 & 15.0793 \\
\hspace{1cm}$(1,1,1,1)$        &11.0414 &10.0325 &15.7677 & 16.8896\\
\hspace{1cm}$(2,1,1,1)$        & 10.4128 & 9.4538 &14.8478 & 15.8523\\

$^V\mathcal{M}^{(0)}_{K+1K+11}$&  4.7566 & 4.3644 &$--$& $--$ \\
\hspace{1cm}$(1,1,1,1)$        &  5.3167 & 4.9067 &$--$& $--$ \\
\hspace{1cm}$(2,1,1,1)$        & 4.9860 & 4.6078  &$--$& $--$\\

$^A\mathcal{M}^{(0)}_{K+1K+11}$&  2.2945 & 1.4936  &$--$& $--$\\
\hspace{1cm}$(1,1,1,1)$        &   2.8225 & 1.8565 &$--$& $--$ \\
\hspace{1cm}$(2,1,1,1)$        &  2.7069 & 1.7847 &$--$& $--$ \\
\end{tabular}
\end{ruledtabular}
\end{table*}

\begin{table*}[!ht]
  \caption{Values of the s-NME
$^V\mathcal{M}_{KK-11}^{(0)}$ for the studied transitions in $^{59,60}$Fe. For these values
the calculations reproduce the measured half-life for each value of $g_{\rm A}$.
}\label{table:nmes_CVC}
\begin{ruledtabular}
\begin{tabular}{lcccc}

   & \multicolumn{2}{c} {\hspace{-0.3cm}$^{59}$Fe$(3/2^-)\to\,^{59}$Co($7/2^-$) } &\multicolumn{2}{c}{$^{60}$Fe$(0^+)\to\,^{60}$Co($2^+)$}  \T\B\\
\cline{2-3}
\cline{4-5}
$g_{\rm A}$ & KB3G & GXPF1A & KB3G & GXPF1A\T\B \\
\hline

0.80 & 0.10371 & 0.09499 &0.11429&0.09187 \T\\
0.90 & 0.09883 & 0.09009&0.10718&0.08093\T\\
1.00 & 0.09391 & 0.08515&0.10003&0.06987\T\\
1.10 & 0.08894 & 0.08017&0.09282&0.05869\T\\
1.20 & 0.08393 & 0.07515&0.08555&0.04739\T\\
1.27 & 0.08039 & 0.07161&0.08044&0.03941\B\\
\end{tabular}
\end{ruledtabular}
\end{table*}

\subsection{Nuclear matrix elements}

Once the shell-model description of low-energy spectra and spectroscopic properties of the 
involved nuclei is now under control, we are ready to use the resulting wave functions to 
compute the OBTDs needed in the NMEs for the $\beta$-decay-rate calculations.

The main theme of the present calculations is to vary the value of the axial coupling strength
$g_{\rm A}$ and see how it affects the electron spectral shapes of the second-forbidden 
nonunique $\beta^-$-decay transitions $^{59}\textrm{Fe}(3/2^-)\to\,^{59}\textrm{Co}(7/2^-)$ and
$^{60}\textrm{Fe}(0^+)\to\,^{60}\textrm{Co}(2^+)$. In order to (possibly) gain a rough idea
about the possible values of this coupling in the presently discussed mass region, we have
studied the decay rates of two measured allowed transitions in $^{59}$Fe.  
In Table \ref{table:mgt}, we present our calculated absolute values of the Gamow-Teller 
($\mathcal{M}_{\rm GT}$) and Fermi ( $\mathcal{M}_{\rm F}$) matrix elements for these allowed 
transitions from the $3/2^-$ ground state of $^{59}$Fe to the first two $3/2^-$ excited 
states in $^{59}$Co. The experimental $\mathcal{M}_{\rm GT}$ NMEs are obtained by using the 
experimental $\log ft$ values \cite{nndc} and the information that the Fermi NMEs can be 
neglected based on our calculated numbers, presented in the table.
Fermi matrix elements for the allowed transitions in $^{59}$Fe are tiny due to isospin conservation. 
For the transition $3/2^-\to\,3/2^-_1$, the KB3G-computed magnitude of $\mathcal{M}_{\rm GT}$ 
is found consistent with the experimental one, while for the GXPF1A interaction the
computed value is far from the experimental one. In the case of the $3/2^-\to\,3/2^-_2$ 
transition, our calculated absolute values of $\mathcal{M}_{\rm GT}$ for both interactions
are far too small compared to the data. Unfortunately, our attempt, did not lead to any 
clarification of the $g_{\rm A}$ problem due to the vastly different predicted values of the
Gamow-Teller matrix elements for the two interactions and two transitions. It seems that 
the prediction of the exact structure of the higher-lying excited states is not so trivial 
for the presently used shell-model interactions. Therefore, we have to perform our SSM analyses 
for the second-forbidden nonunique $\beta^-$ decays of $^{59,60}$Fe without any perception of
a possible preferred effective value of $g_{\rm A}$. 

The KB3G- and GXPF1A-computed NMEs for the second-forbidden nonunique $\beta^-$ decays of 
$^{59,60}$Fe are presented in Table \ref{table:nmes}. The relativistic vector NME 
$^V\mathcal{M}^{(0)}_{KK-11}$ becomes identically zero due to the limitation of our adopted 
single-particle model space (see more details in Refs. \cite{Anil2020PRC, Kirsebom2019PRC}). 
For convenience of notation, we will call this NME as s-NME (small NME).
As mentioned in the introduction, we constrain the value of this matrix element by two 
different ways: either i) from the CVC relation as described in \cite{behrens1982}, or 
ii) by reproducing the experimental partial half-life by tuning this matrix element separately
for each value of $g_{\rm A}$. In the method i), the mentioned CVC relation \cite{behrens1982} 
reads

\begin{eqnarray}\label{eqn:CVC}
^VF_{211}^{(0)}=-\frac{1}{\sqrt{10}}\left[W_0R-(M_n-M_p)R+
\frac{6}{5}\alpha{Z}\right]{^VF_{220}^{(0)}}, \nonumber\\ 
\end{eqnarray}
where the $^VF_{KLS}^{(N)}$ are the form factor coefficient which are related to the NMEs via

\begin{eqnarray}
R^L{^VF_{KLS}^{(N)}}(k_e,m,n,\rho)=(-1)^{K-L}g_{\rm V}{^V\mathcal{M}}_{KLS}^{(N)}(k_e,m,n,\rho). 
\nonumber\\
\end{eqnarray}

The values of the CVC-constrained small nuclear matrix element s-NME, obtained from the above 
relations using the computed values of the large vector NME $^V\mathcal{M}^{(0)}_{KK0}$, 
are presented in Table \ref{table:nmes} with the additional label ``CVC''. The large NME
receives the bulk of its contributions from the presently used single-particle model space so
that its value can be considered as quite reasonable.
In the method ii), we have used the s-NME as a fit parameter to reproduce the measured 
partial half-life of the decay transition for each $g_{\rm A}$ separately, and the 
corresponding s-NME values are presented in Table \ref{table:nmes_CVC}.  In Table 
\ref{table:nmes}, the CVC-constrained s-NMEs are compared with the fit-parameter s-NMEs, 
obtained by using $g_{\rm A}=0.80$. 
 It can be seen that the fitted values of 
the s-NMEs are of the same order of magnitude as the CVC-constrained ones, differing by a
factor $2-3$ from each other. 

As seen in Tables \ref{table:nmes} and \ref{table:nmes_CVC}, the leading-order (LO) s-NME, 
$^V\mathcal{M}^{(0)}_{KK-11}$, is roughly three orders of magnitude smaller than the LO 
large vector NME $^V\mathcal{M}^{(0)}_{KK0}$. However, owing to the  systematic order-by-order 
expansion of Behrens and B\"uhring \cite{behrens1982}, their contributions to the 
$M_K(k_e,k_\nu)$ and $m_K(k_e,k_\nu)$ matrix elements of Eq. (\ref{eq2}) are on equal footing. 
This is quaranteed by the smaller phase-space factors multiplying the 
$^V\mathcal{M}^{(0)}_{KK0}$ matrix element. A similar argument can be used when comparing 
the relative magnitudes of the matrix elements in the NLO. The contribution of the NLO 
relative to that of the LO, is suppressed by the overall small phase-space factors, 
as discussed in  \cite{mika2017,behrens1982}.

\begin{figure*}[ht]
\centering
\includegraphics[width=\columnwidth]{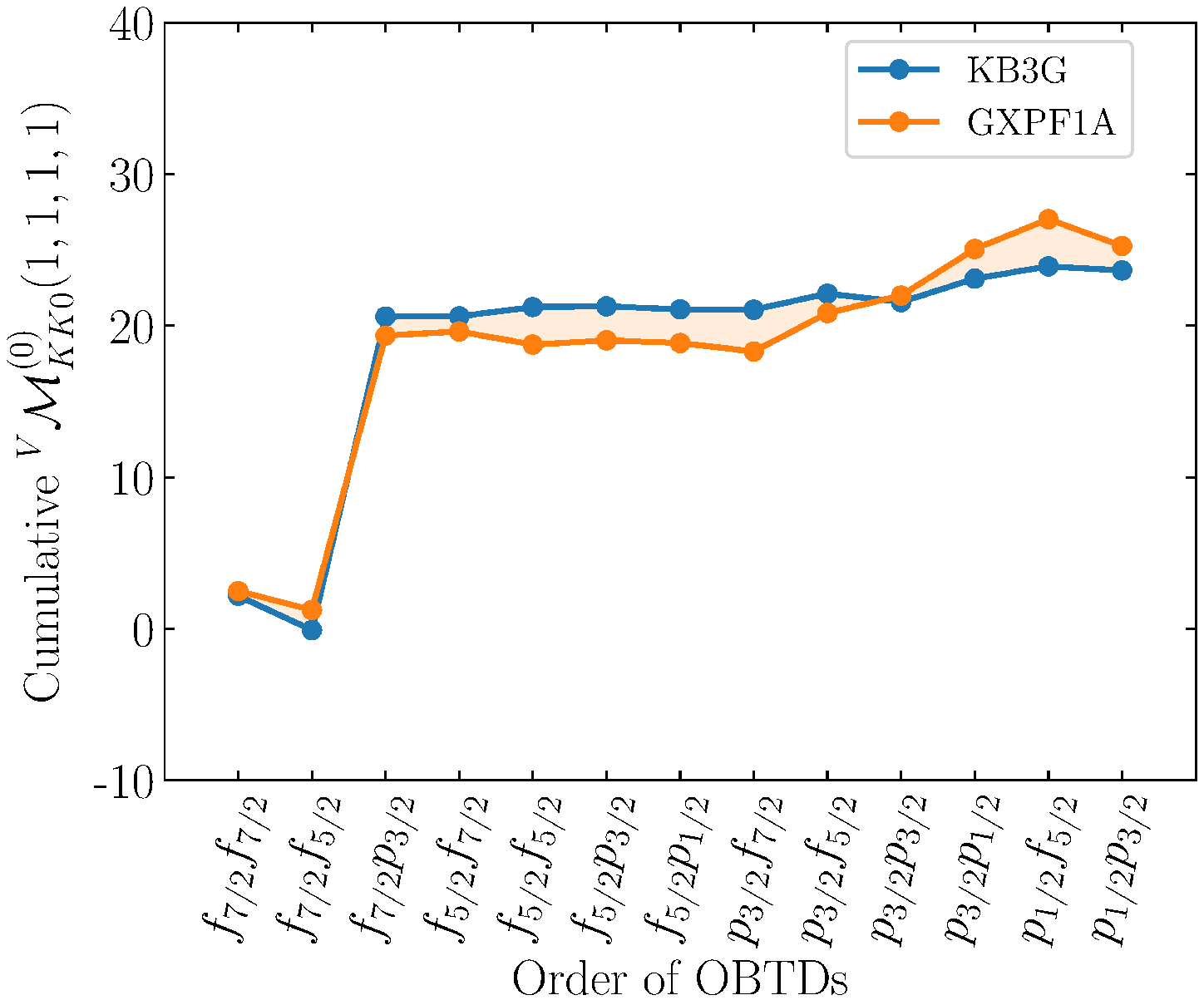}
\includegraphics[width=\columnwidth]{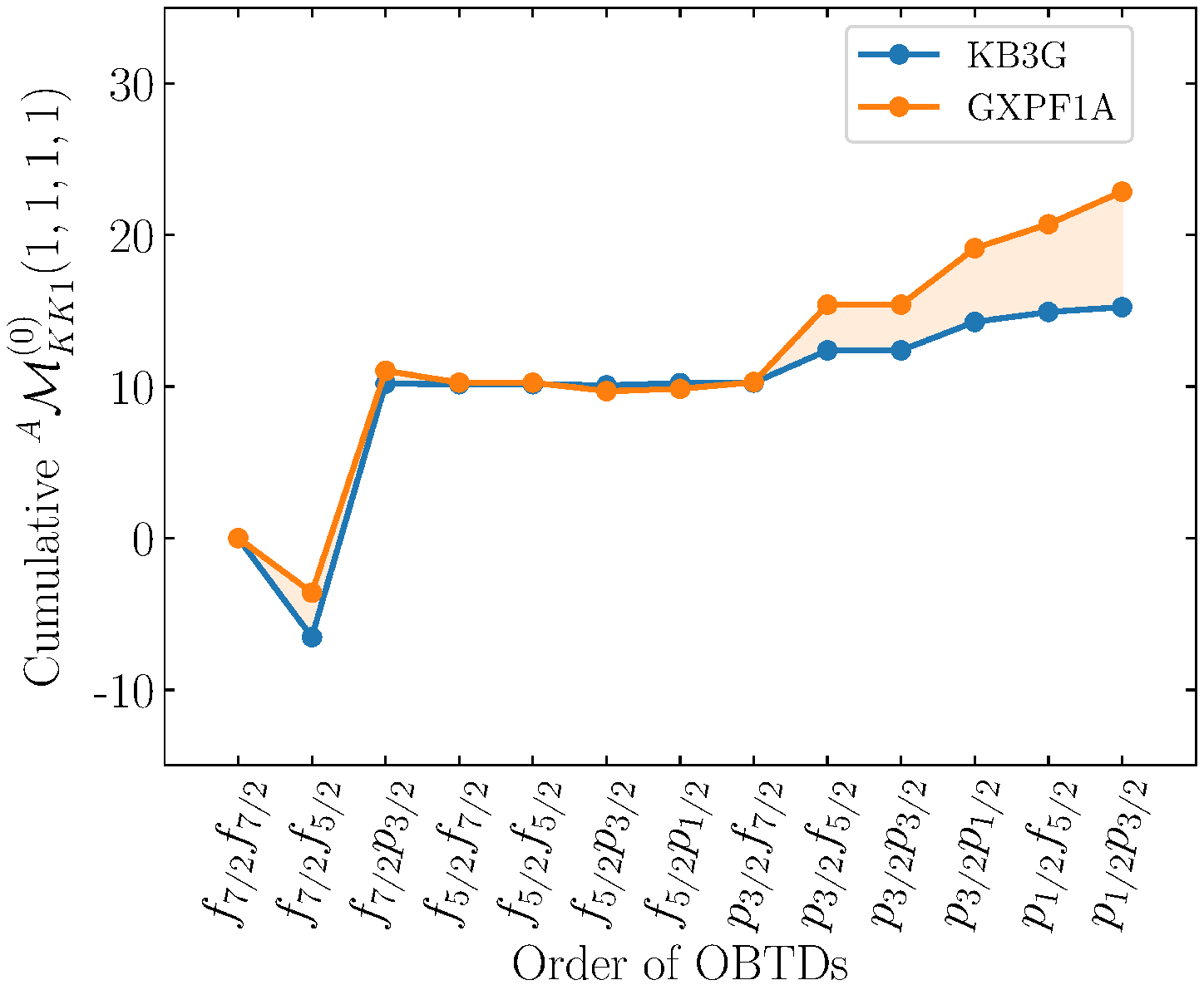}
\caption{Cumulative sum of the vector matrix element 
$^V\mathcal{M}_{KK0}^{(0)}$(1,1,1,1) (left panel) and the axial-vector matrix element
$^A\mathcal{M}_{KK1}^{(0)}$(1,1,1,1) (right panel) for the decay of $^{60}$Fe. The horizontal 
axis lists the contributing proton-neutron orbital pairs.}\label{fig:60Fe_nmes}
\end{figure*}

As seen in Table \ref{table:nmes}, the NMEs for $^{59}$Fe have consistent values for the two 
interactions. In the case of $^{60}$Fe, the magnitudes of the computed axial-vector matrix 
elements are found to be larger for the GXPF1A interaction 
than for the KB3G interaction, leading to strong differences in the shape factors.
This difference can be traced back to the differences in the behavior of the
cumulative sums of the vector and axial-vector matrix elements, exemplified by the 
vector matrix element $^V\mathcal{M}_{KK0}^{(0)}$(1,1,1,1) and 
the axial-vector matrix element $^A\mathcal{M}_{KK1}^{(0)}$(1,1,1,1) in Fig. \ref{fig:60Fe_nmes}.
In this figure the cumulative sums of these matrix elements are plotted as functions
of the contributing proton-neutron orbitals. As seen in the figure, for the vector matrix
element there is a shift in the contributions of the proton-neutron orbital pairs 
from the very beginning such that these shifts conspire to produce a similar total vector 
matrix element for both interactions. For the axial-vector matrix elements this shift
realizes only for the $p$-$f$ proton-neutron orbital contributions.
The difference between the shell model interactions stems from the contributions of $p$-orbital protons with neutrons not in the $f_{7/2}$ orbital.  This difference in the cumulative behavior is driven by the differences in the values of the vector-type and
axial-vector-type single-particle transition matrix elements $^{V/A}m_{KLS}^{(N)}(pn)$ in 
Eq. (\ref{eq:ME}). Next we use these computed NMEs in order to evaluate the electron 
spectral shapes as functions of $g_{\rm A}$ for the studied transitions.

\begin{figure*}[!ht]
\centering
\includegraphics[width=\columnwidth]{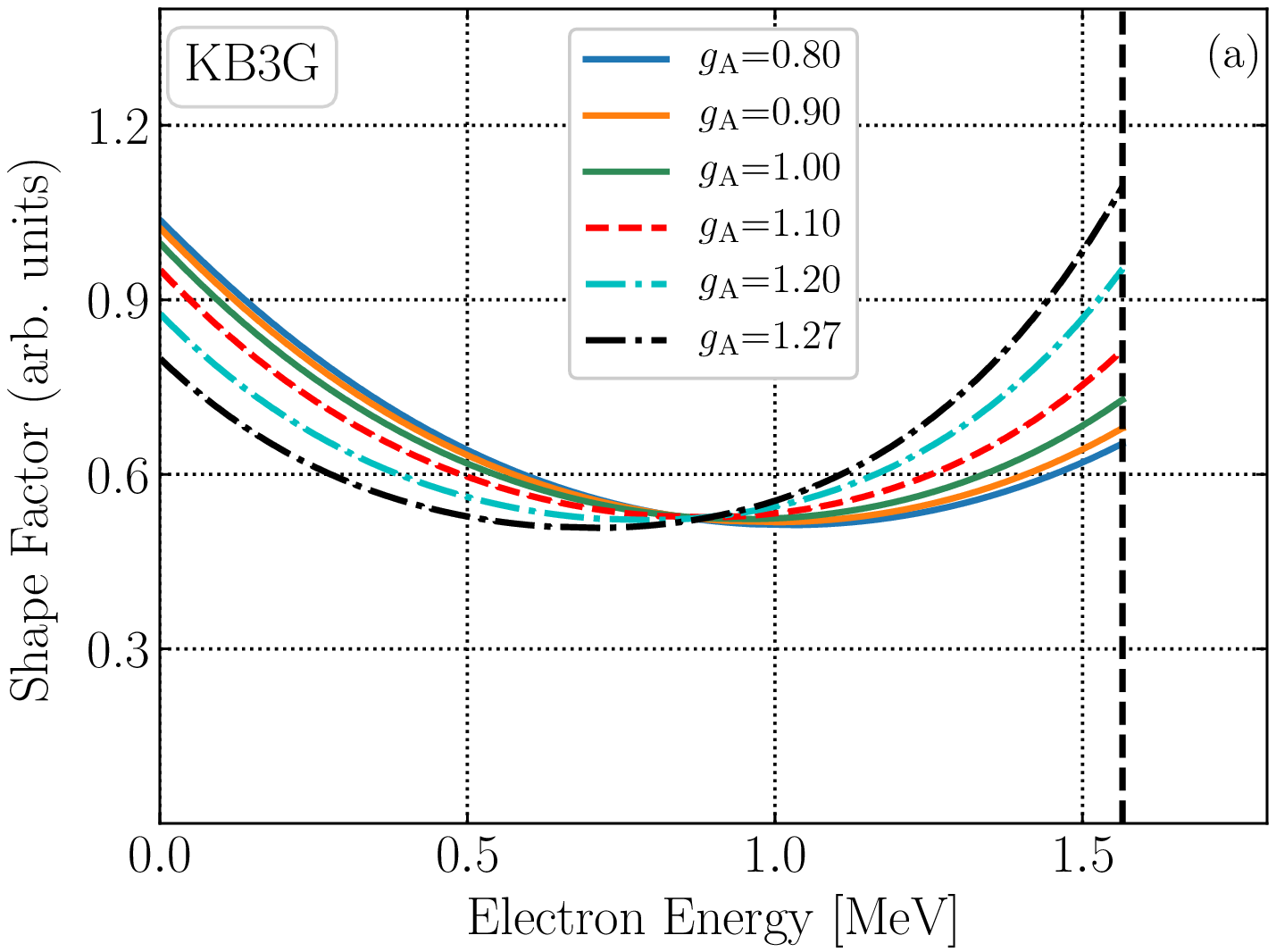}
\includegraphics[width=\columnwidth]{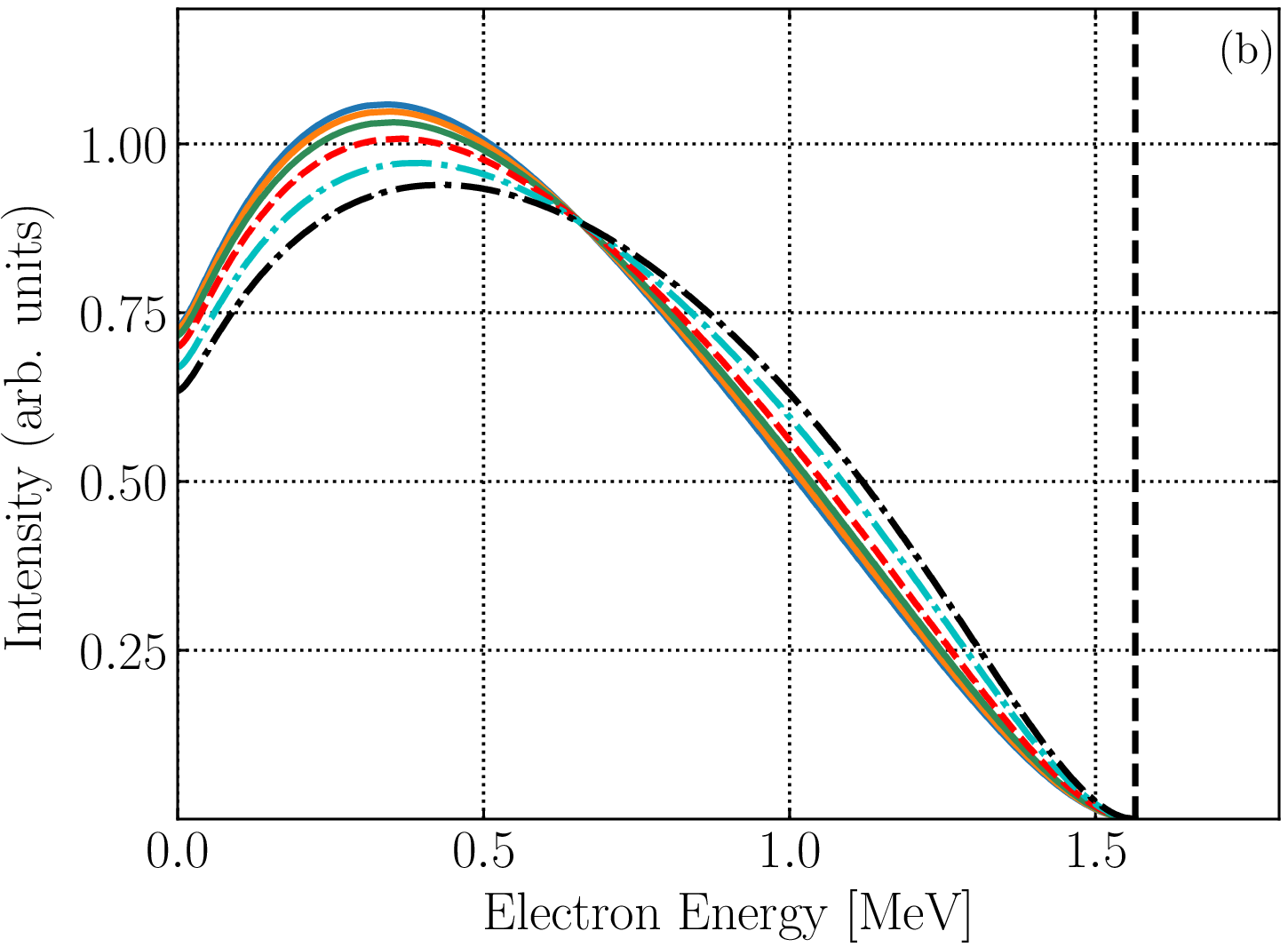}
\includegraphics[width=\columnwidth]{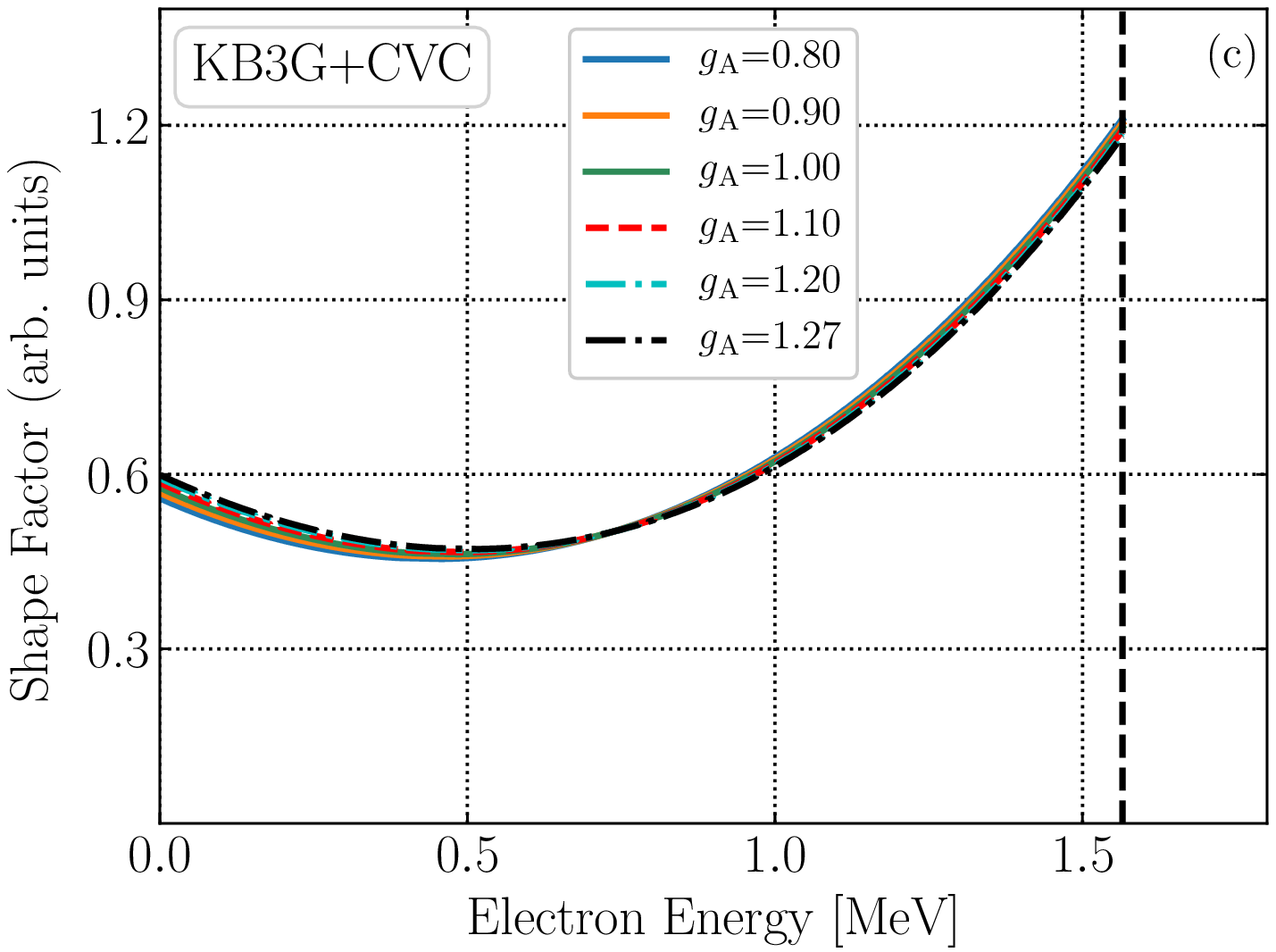}
\includegraphics[width=\columnwidth]{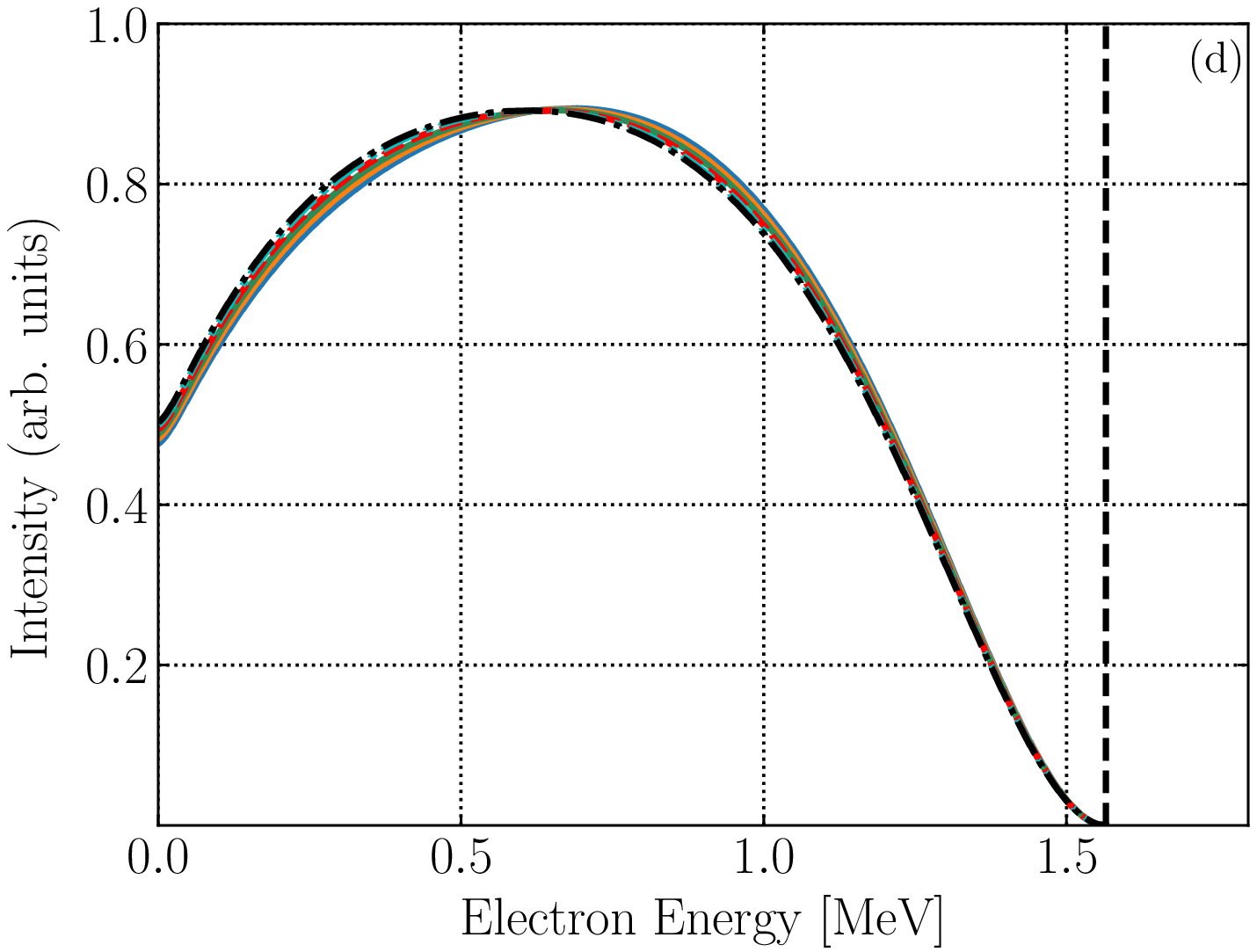}
\includegraphics[width=\columnwidth]{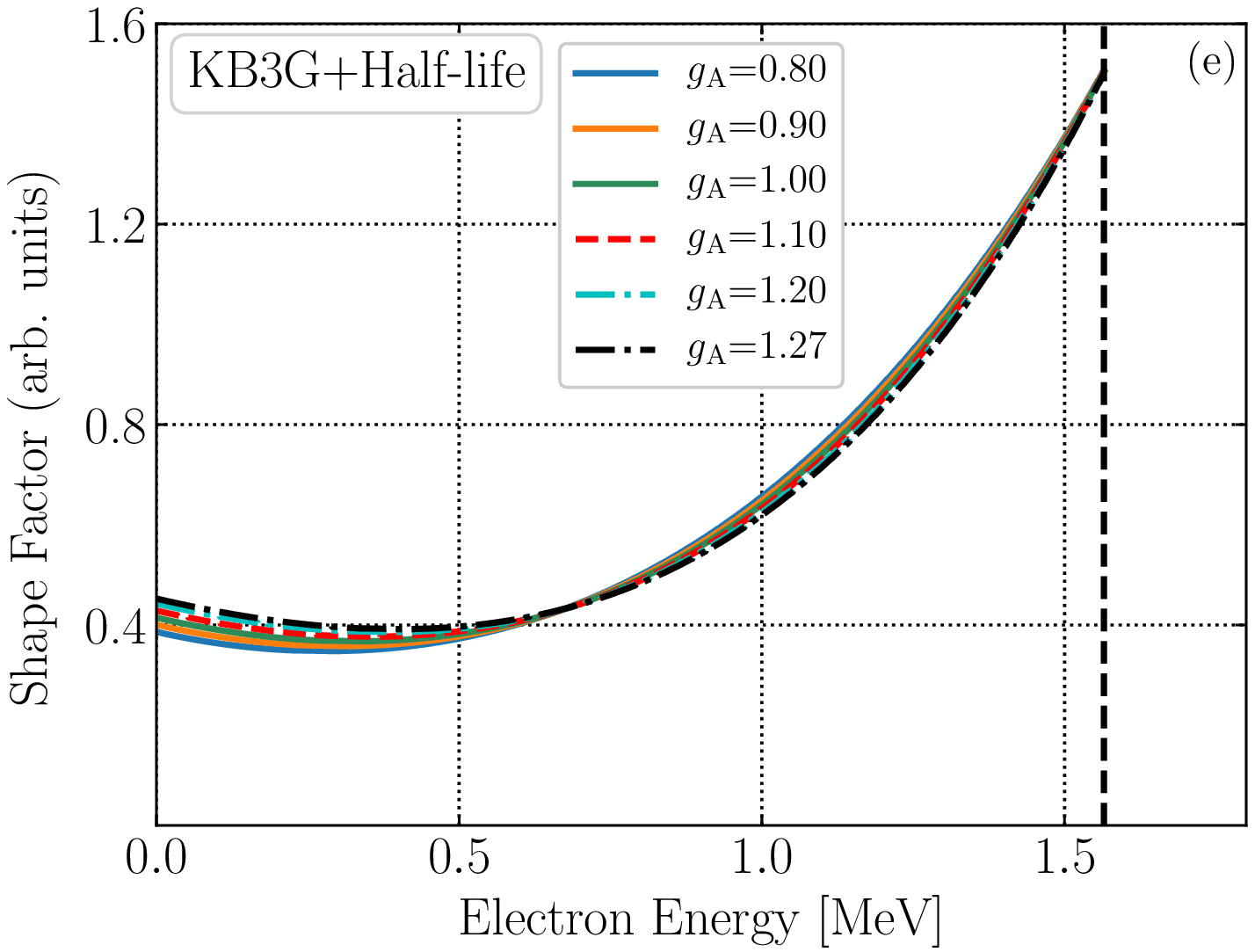}
\includegraphics[width=\columnwidth]{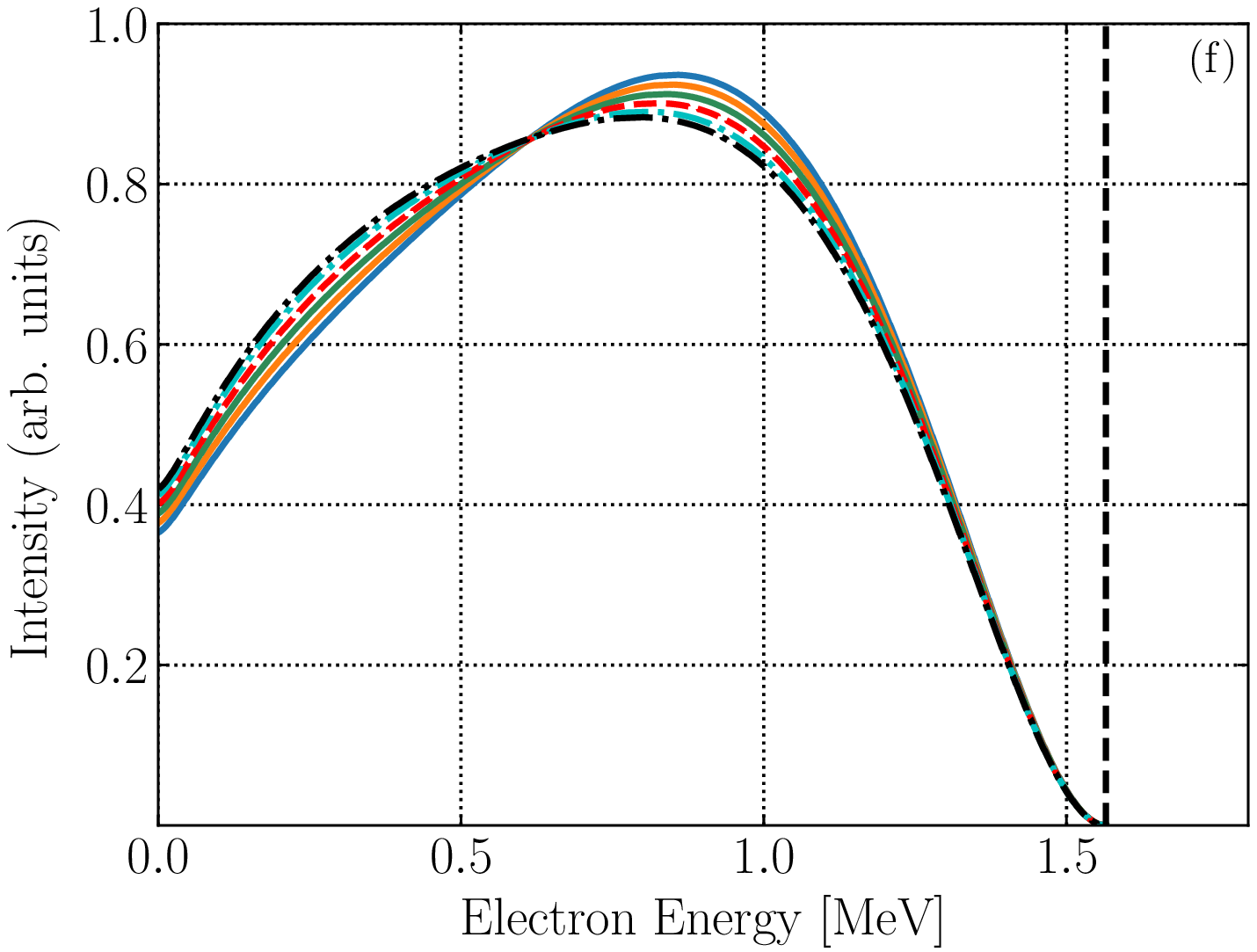}
\caption{The KB3G-computed shape factors (left panels) and electron spectra (right panels) 
as functions of the electron kinetic energy for the second-forbidden nonunique $\beta^-$ 
transition $^{59}$Fe$(3/2^-)\to\,^{59}$Co($7/2^-$). The dashed vertical lines represent the 
end-point energy of the transition and the area under each curve is normalized to unity.
The top panels refer to the results of the original interaction (zero value of the s-NME), 
the figures of the
middle panel use the CVC-constrained value of the s-NME and the figures of the bottom
panel are obtained by using the fitted values of the s-NME.}\label{fig:59Fe_spectra_KB3G}
\end{figure*}

\begin{figure*}[!ht]
\centering
\includegraphics[width=\columnwidth]{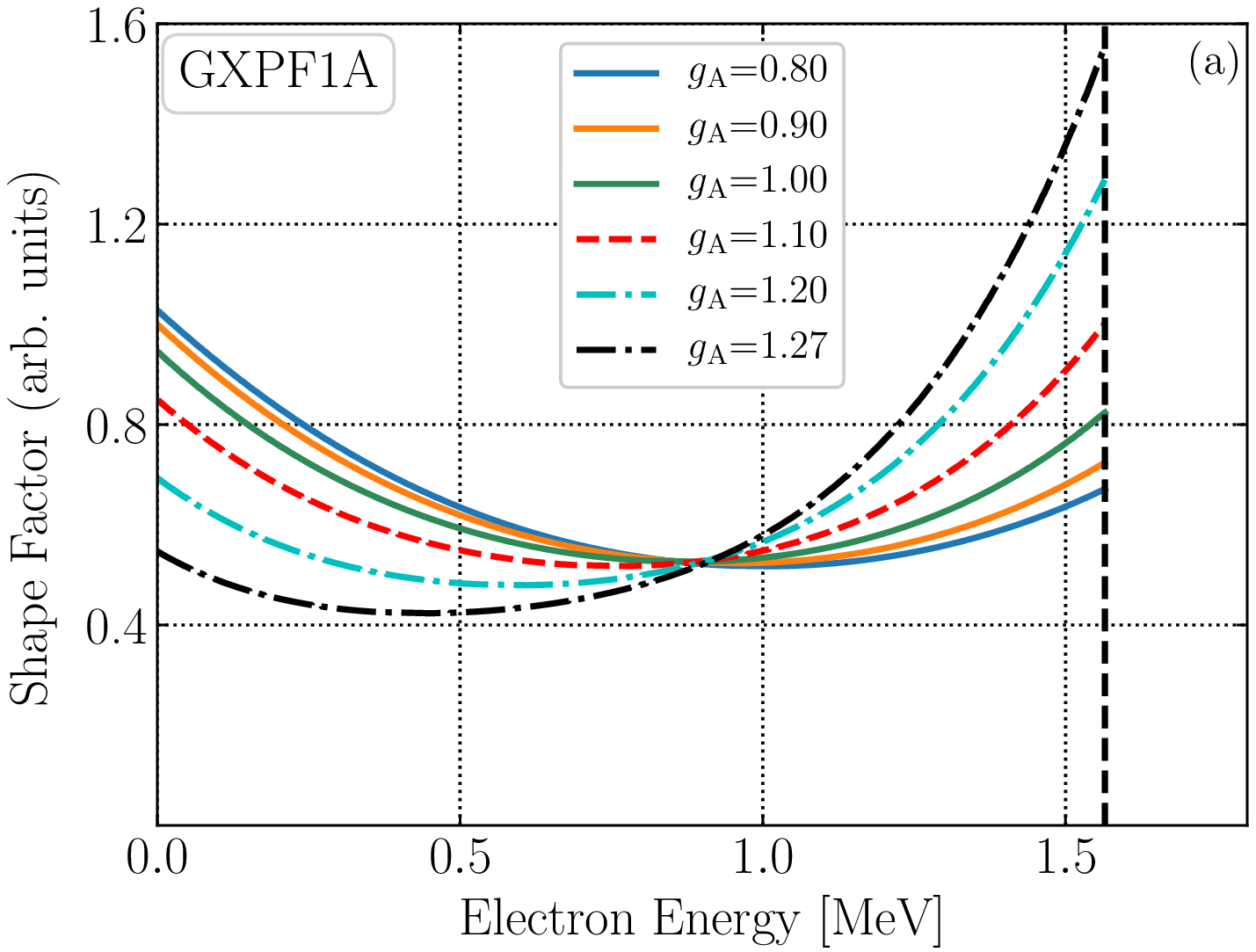}
\includegraphics[width=\columnwidth]{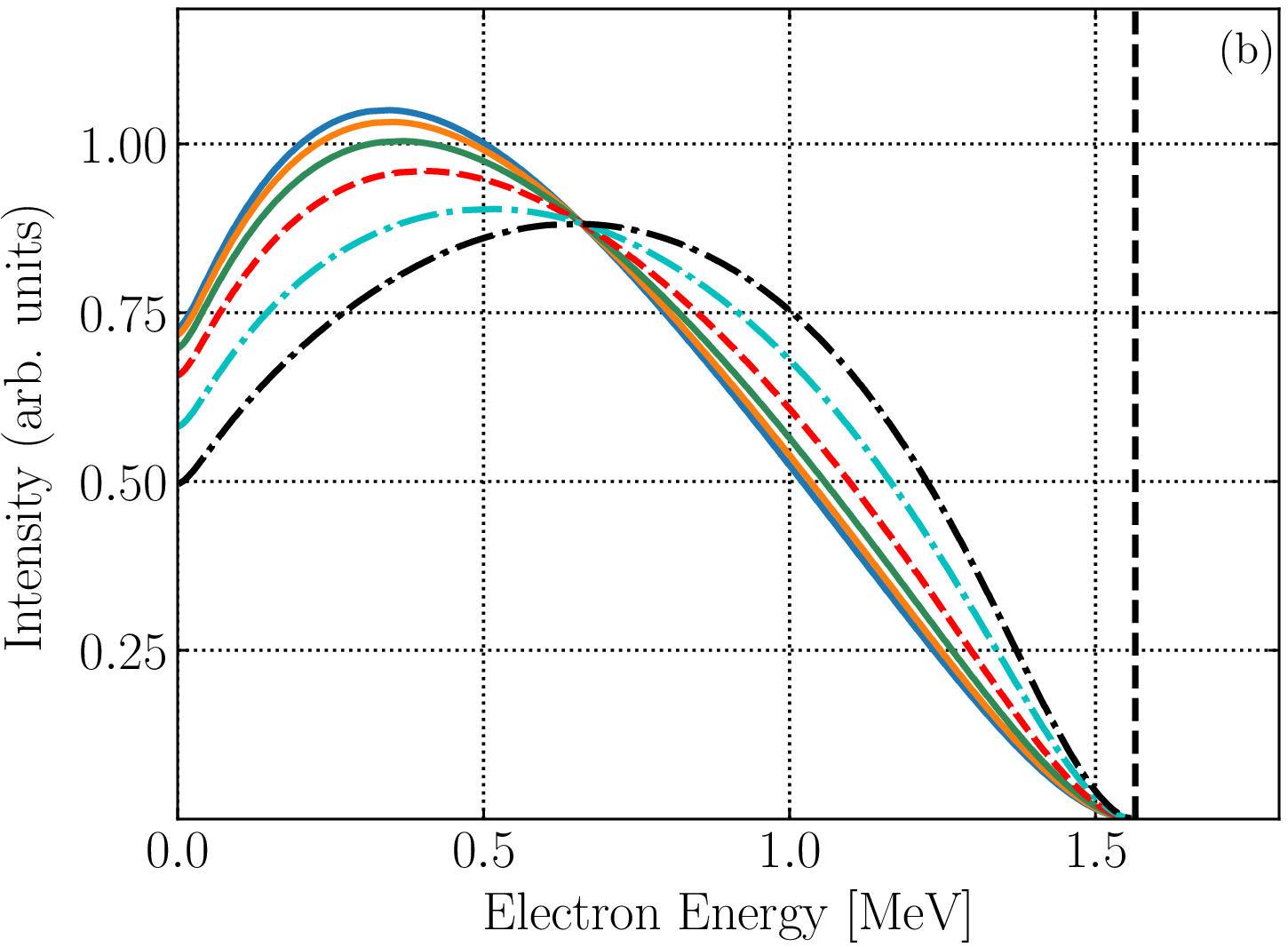}
\includegraphics[width=\columnwidth]{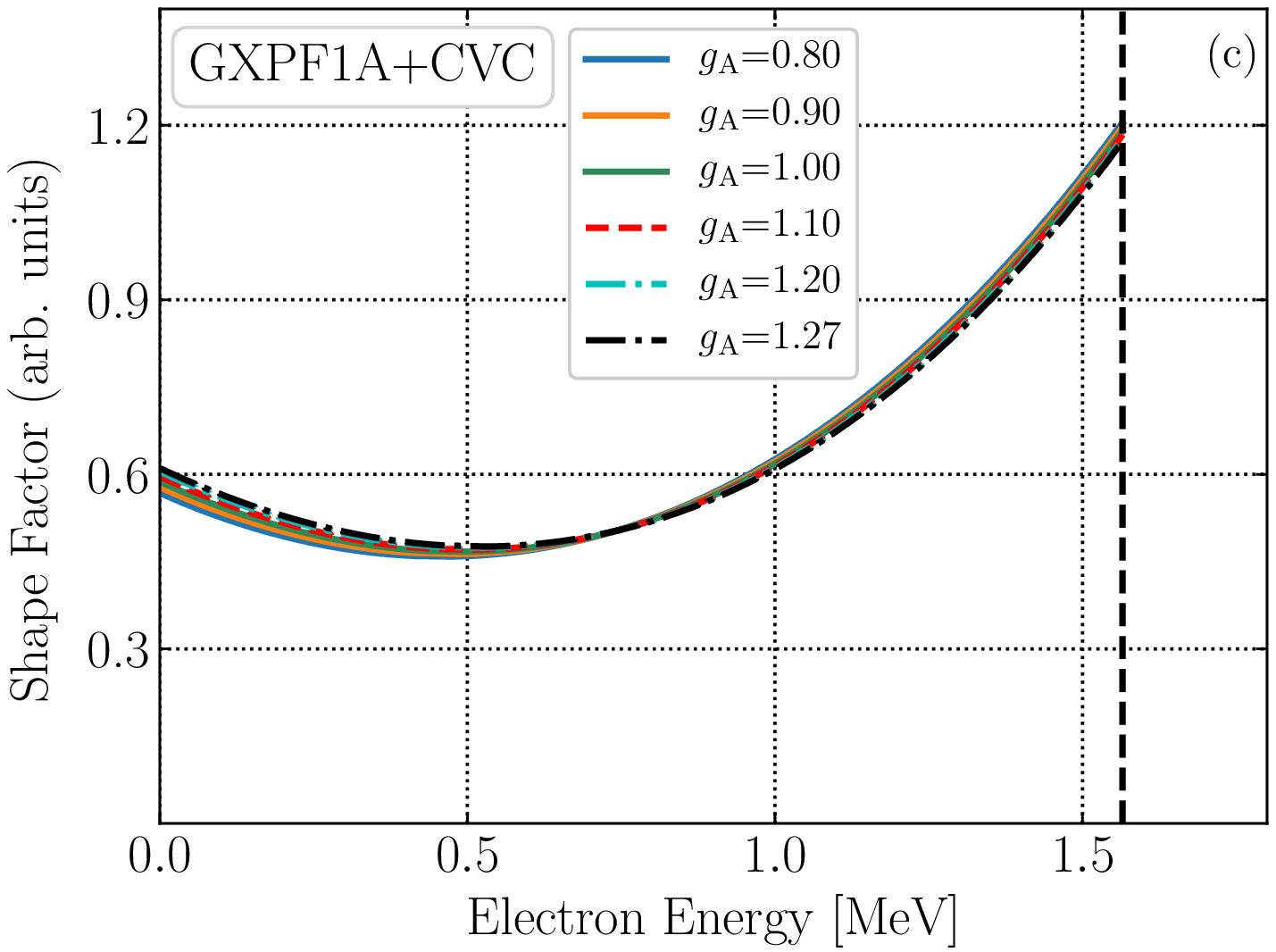}
\includegraphics[width=\columnwidth]{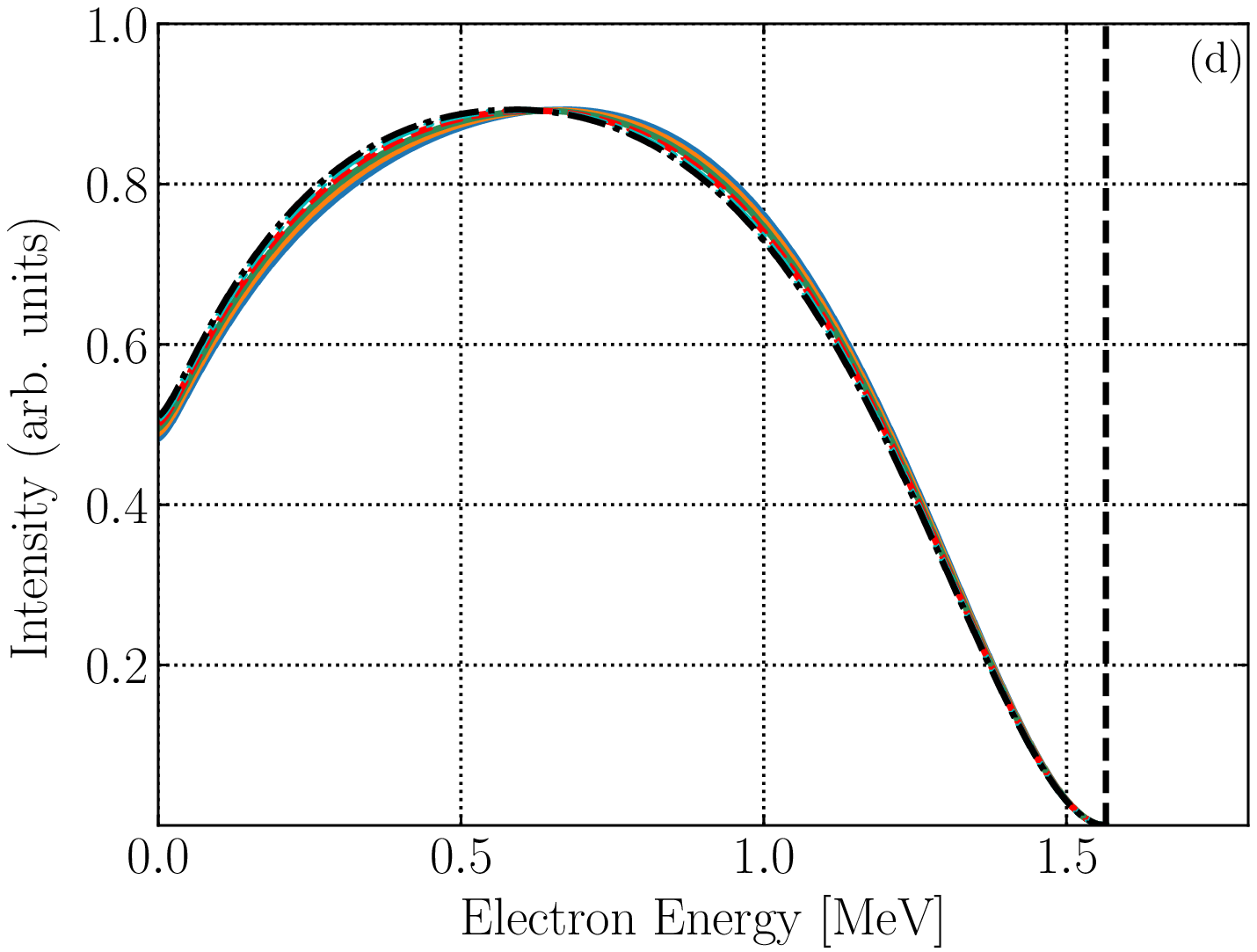}
\includegraphics[width=\columnwidth]{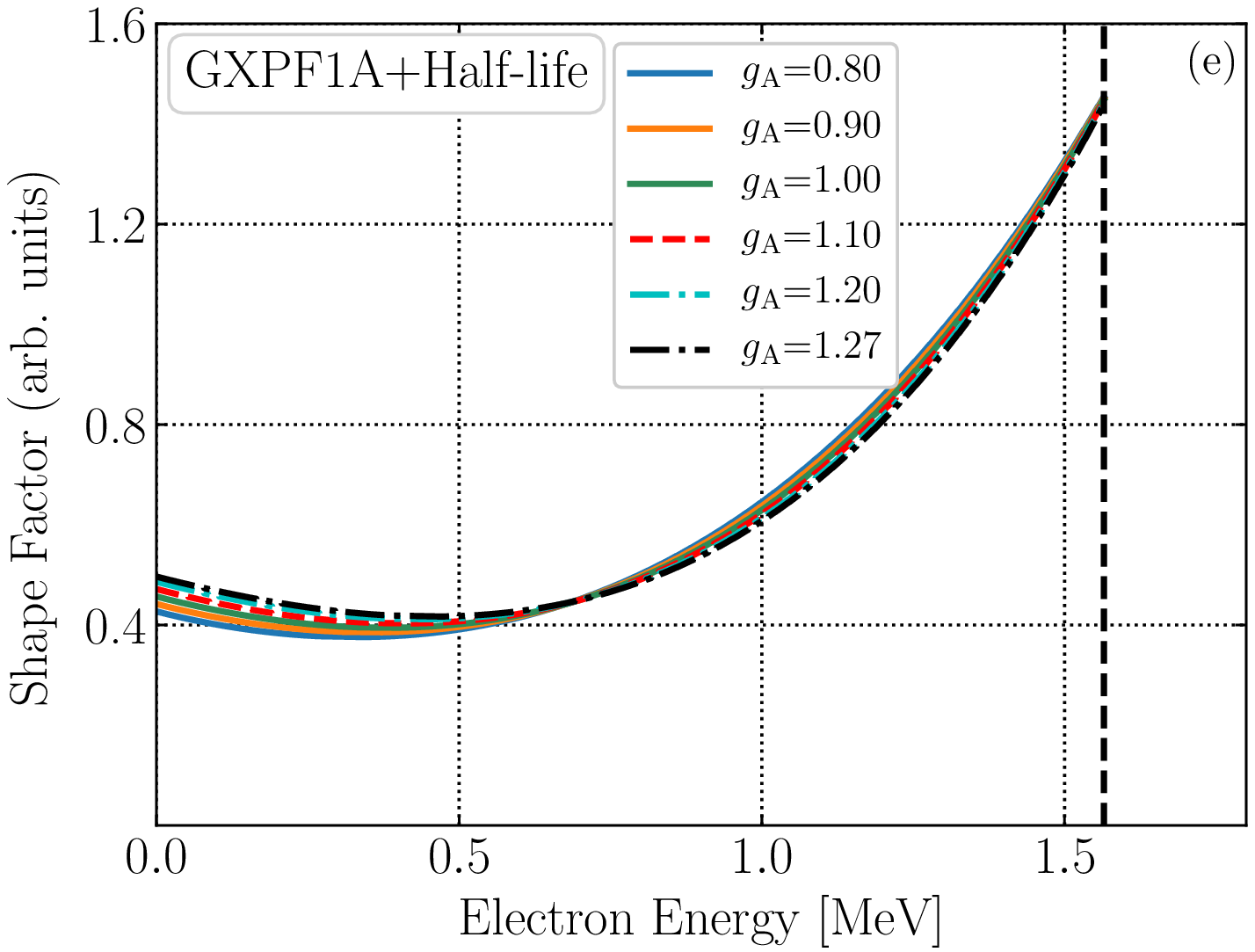}
\includegraphics[width=\columnwidth]{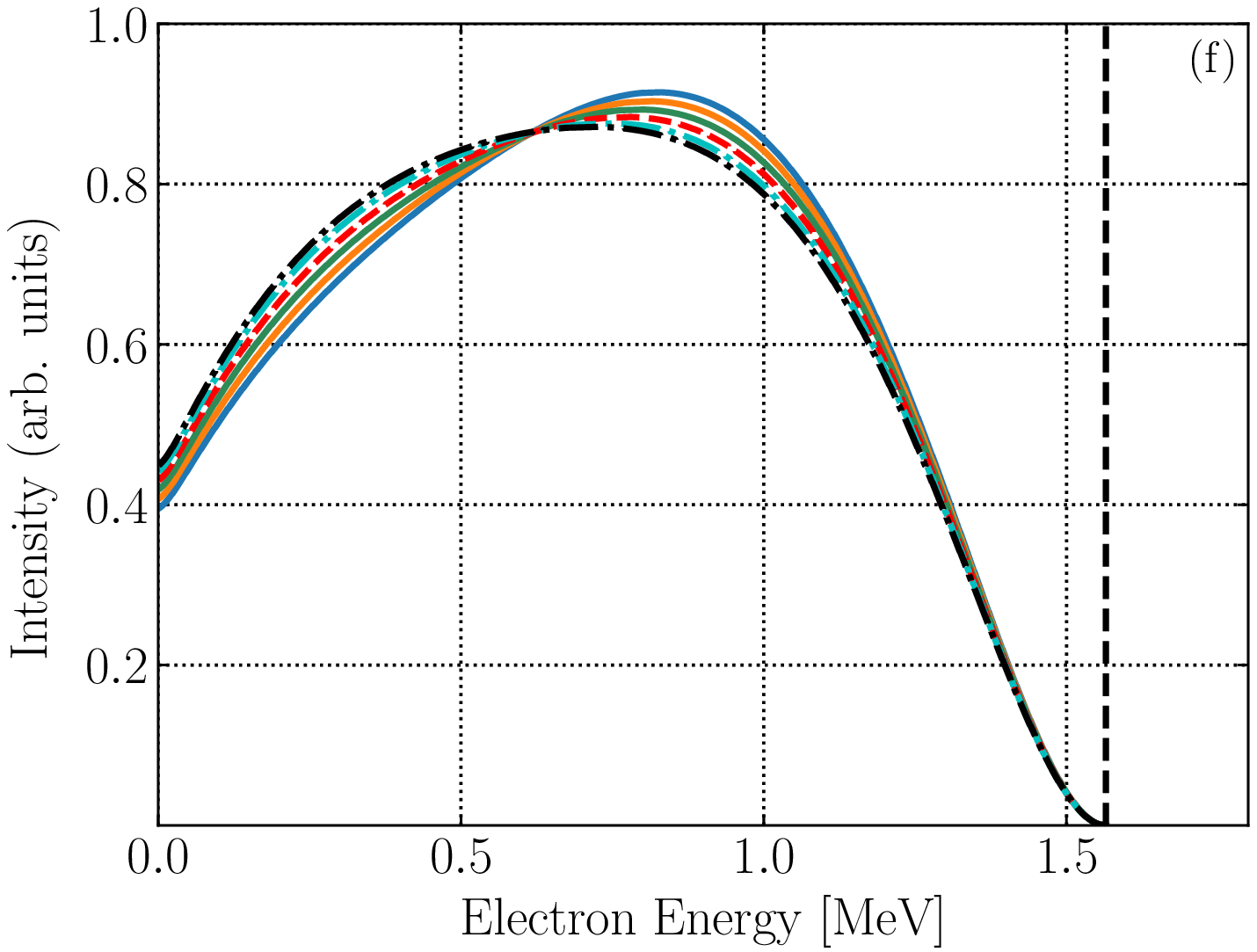}
\caption{The same as in Fig. \ref{fig:59Fe_spectra_KB3G} for the GXPF1A interaction.}
\label{fig:59Fe_spectra_GXPF1A}
\end{figure*}

\subsection{Electron spectral shapes and the effective value of $g_{\rm A}$}

In the SSM, the shape of the $\beta$ spectrum could be used to extract the effective values 
of the weak coupling constants by comparing the computed spectrum with the measured one for 
forbidden nonunique $\beta$ decays. In order to facilitate this comparison with the potentially
measured future spectrum-shape data, we present here the shapes of the electron spectra of the 
second-forbidden nonunique $\beta^-$ decays of $^{59,60}$Fe by varying the value of the 
axial-vector coupling $g_{\rm A}$ within a physically relevant interval. For the electron 
spectral shapes we have performed three sets of calculations in order to see the effects of the 
adopted values of the s-NME on the electron spectral shapes. 
In these calculations the s-NMEs are taken 
0) to come from the bare interaction (thus having zero value), labeled in the figures with 
``interaction name;'' or from the enhanced SSM procedures where the value of the s-NME
comes i) from the CVC relation, labeled in the figures with ``interaction name+CVC'', 
or ii) from the fitting of the measured partial half-life, labeled in the figures with  
``interaction name+Half-life''. The method i) makes the tacit assumption of an ``ideal'' 
calculation, in an infinite single-particle space with all two- and higher-body correlations
included. Method ii) is more appropriate for the present situation where impulse approximation
is assumed, together with severe restrictions in the single-particle space. Also, experimental 
data is best reproduced by using the method ii), so it can be seen as the most reliable in 
predicting the correlation between the spectral shape and the value of $g_{\rm A}$. 
Thus, from the point of view of the application of the
SSM, the $g_{\rm A}$ sensitivity of the spectral shapes obtained by using the method ii) is the
key issue.
It is also to be noted that all the transitions have been observed experimentally but the 
electron spectra are not yet available. For the $\beta^-$-decay transition in $^{59}$Fe the 
measured branching is less than $1\%$ and for the transition in $^{60}$Fe we have a 
$100\%$ branching.

\begin{figure*}[!ht]
\centering
\includegraphics[width=\columnwidth]{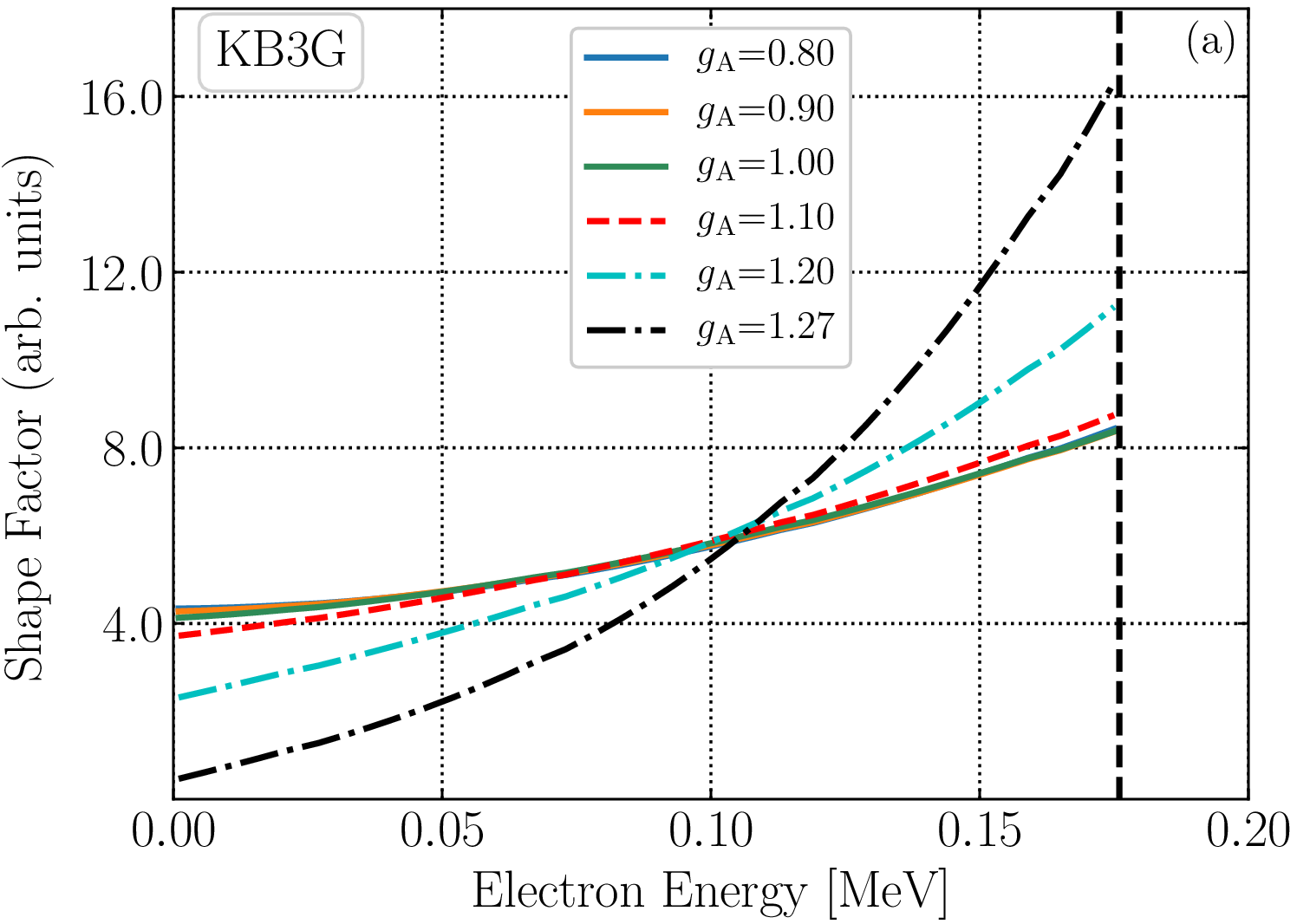}
\includegraphics[width=\columnwidth]{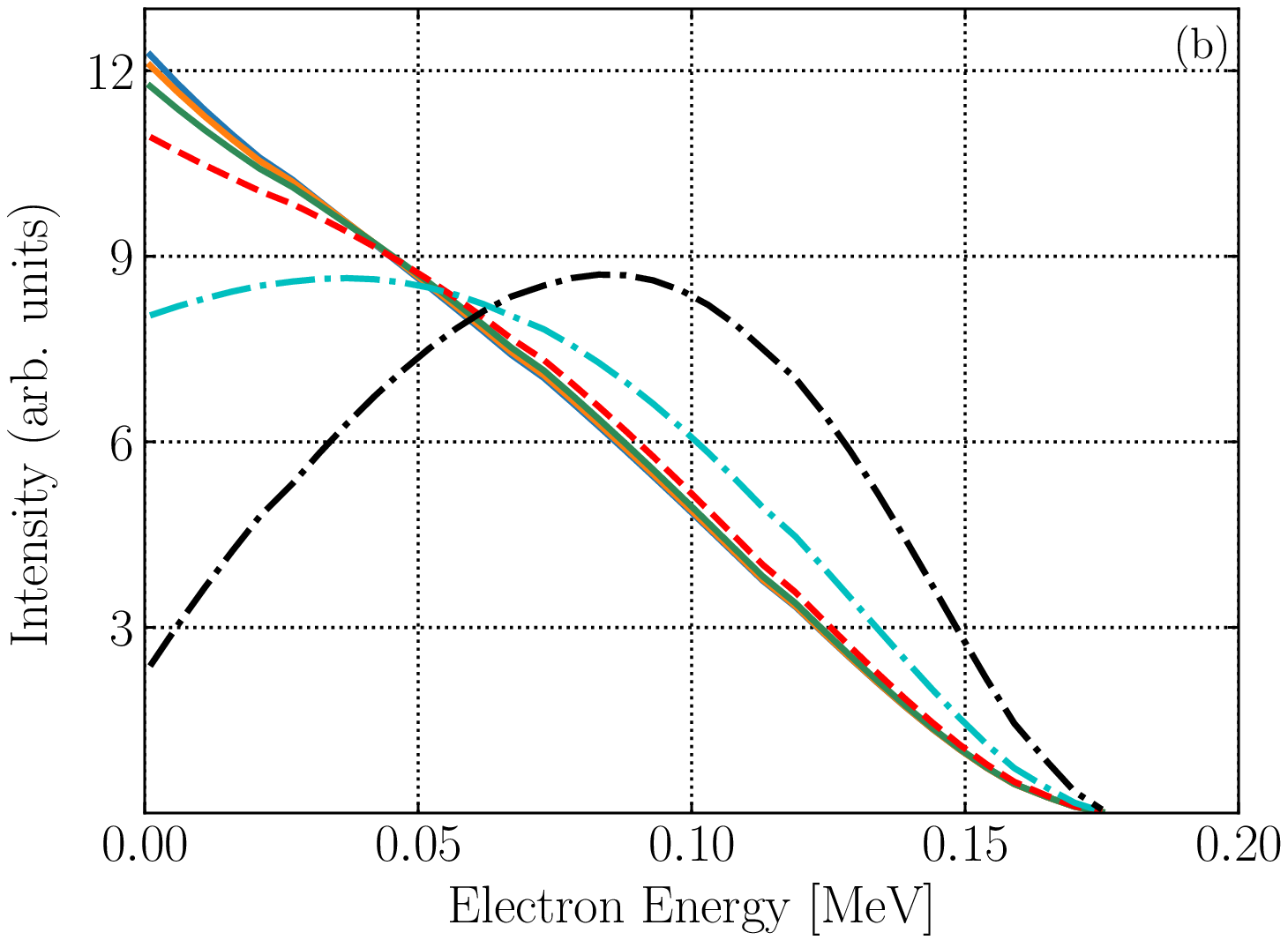}
\includegraphics[width=\columnwidth]{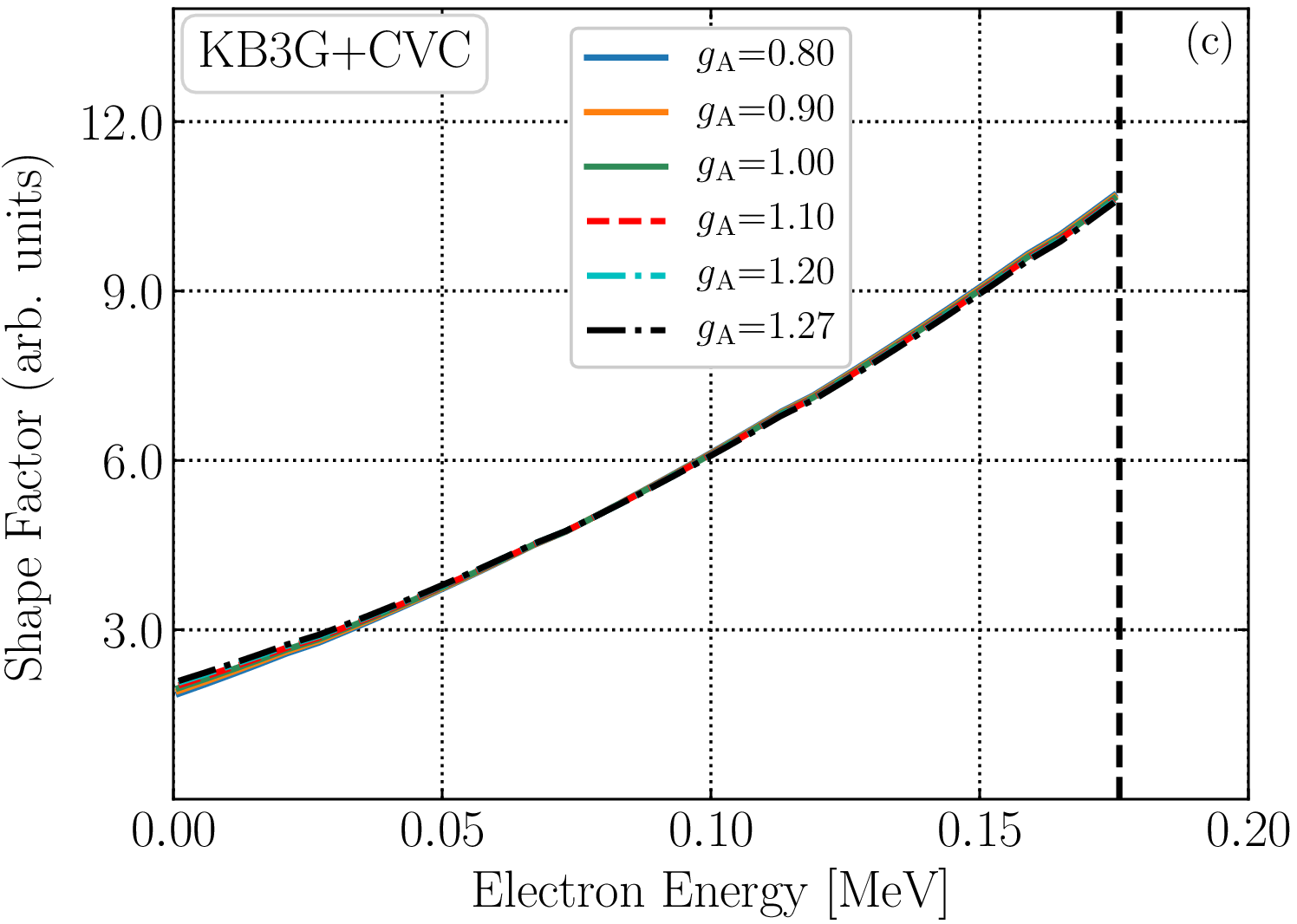}
\includegraphics[width=\columnwidth]{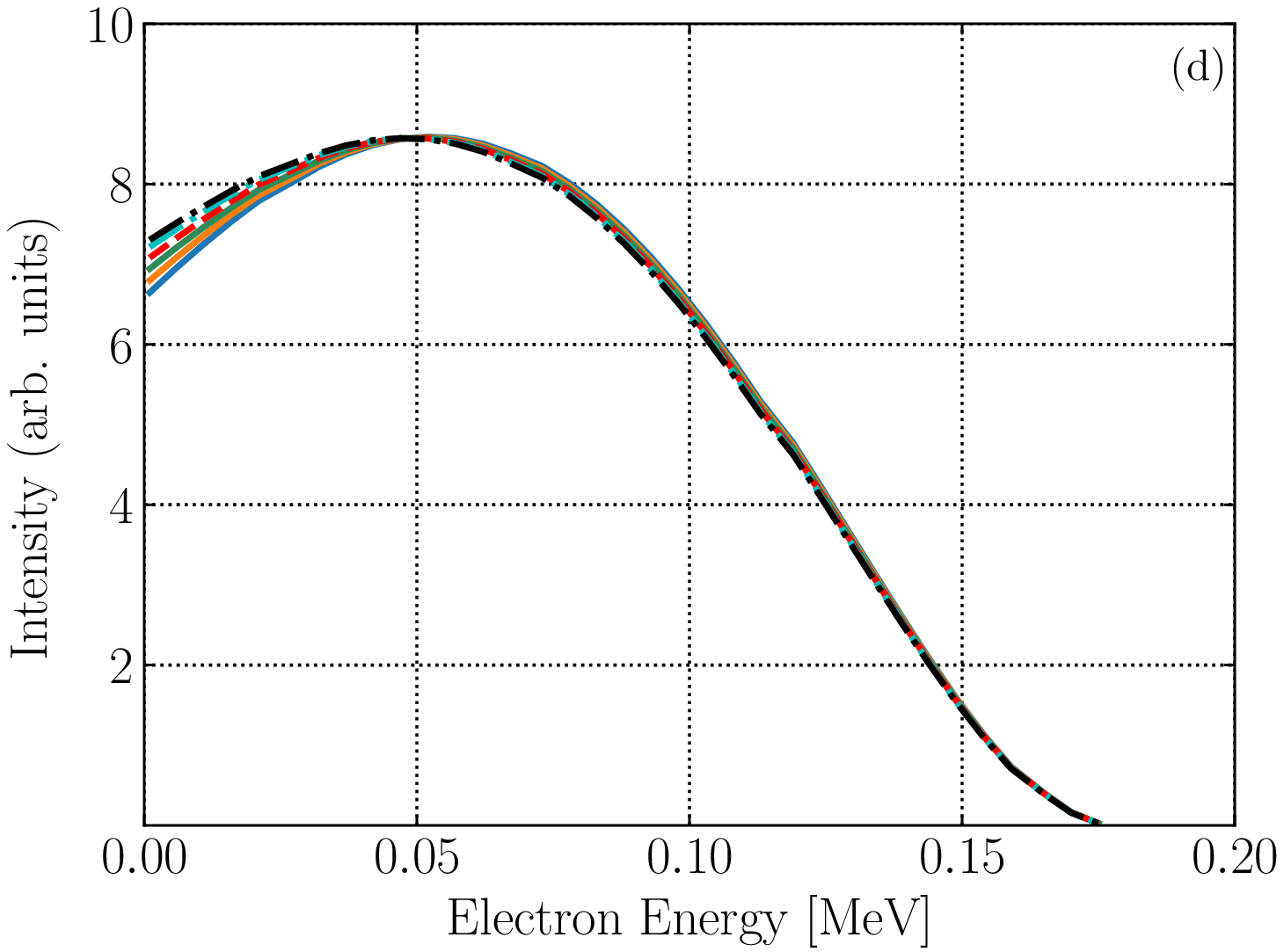}
\includegraphics[width=\columnwidth]{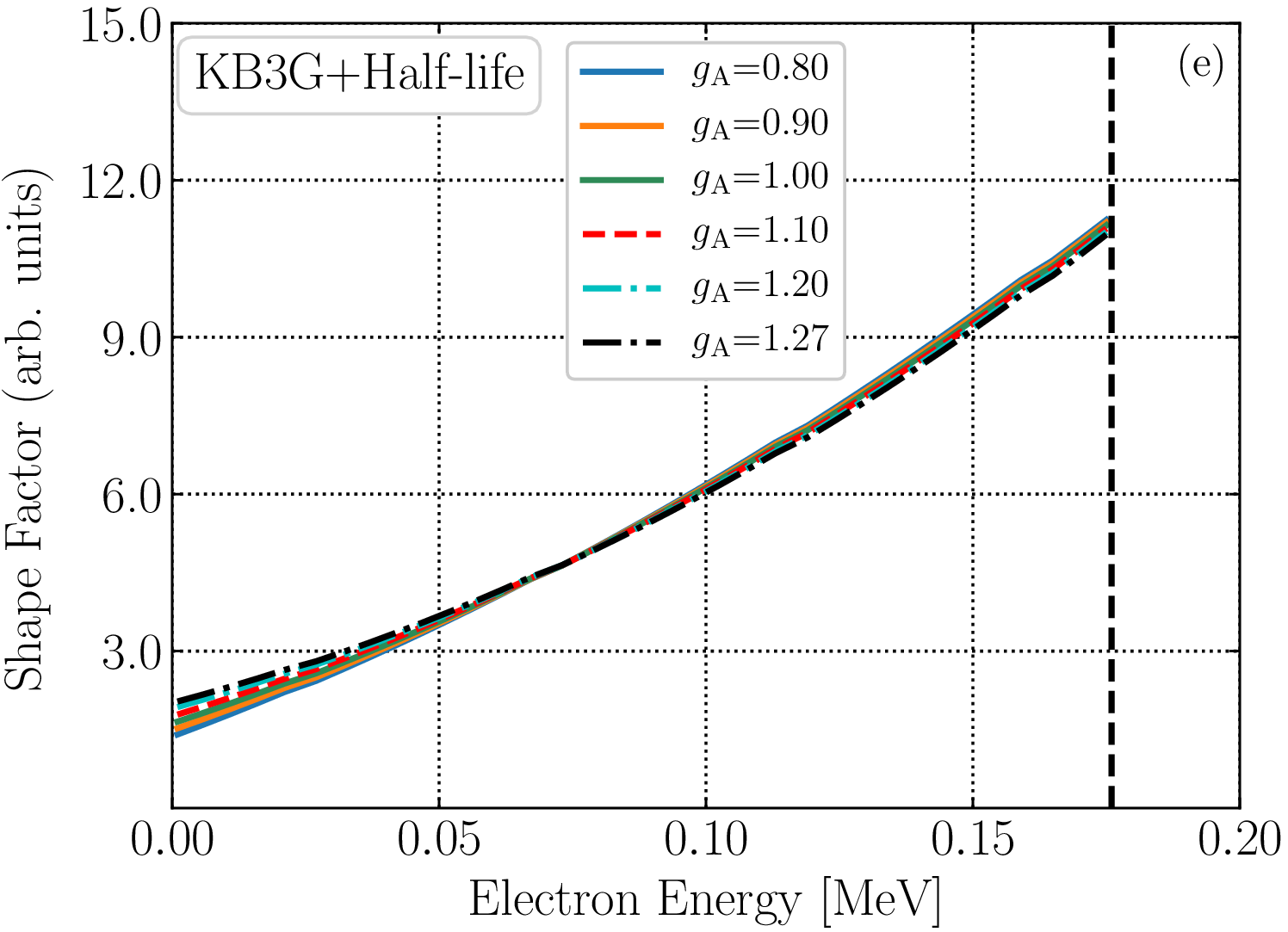}
\includegraphics[width=\columnwidth]{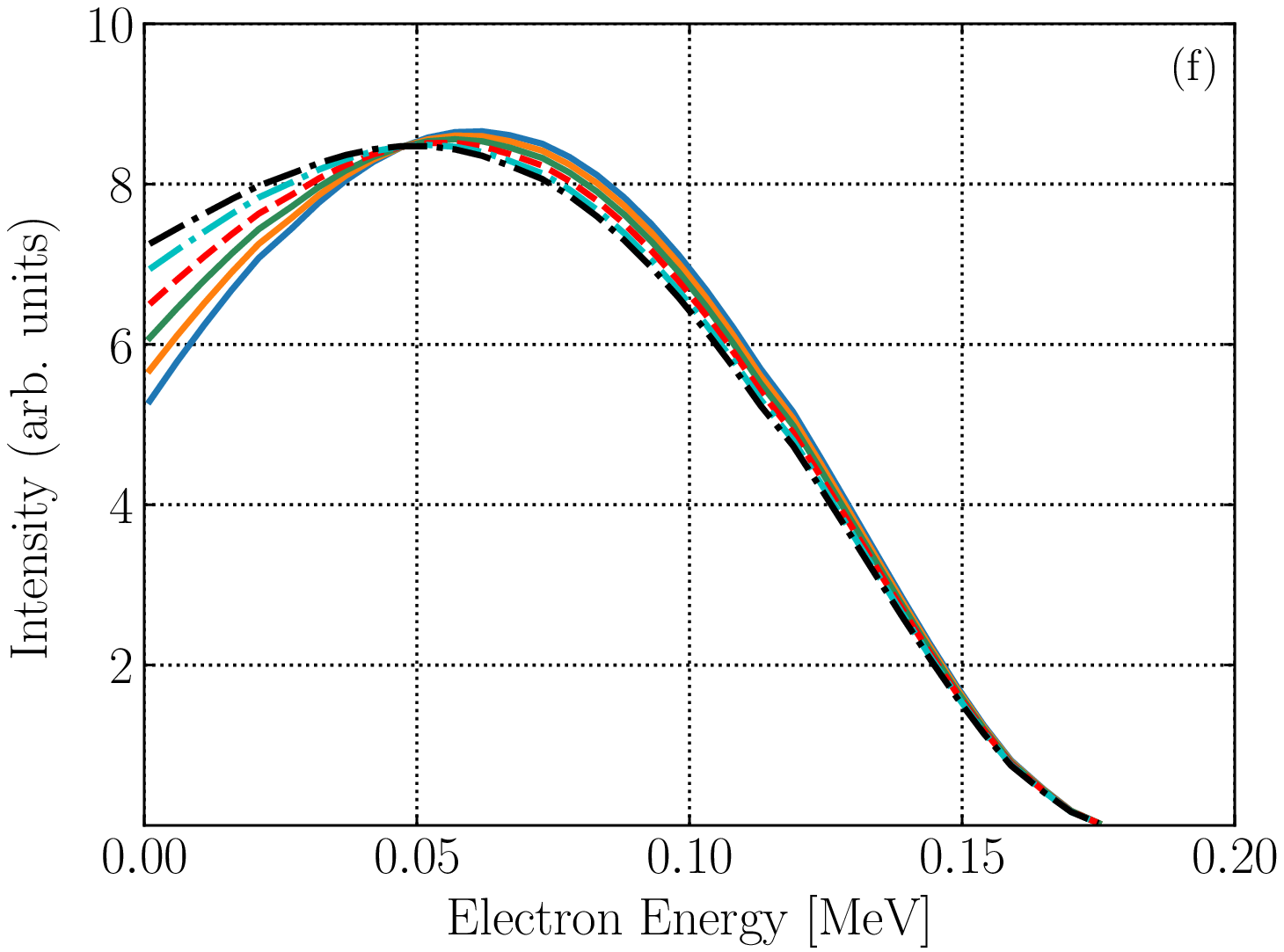}
\caption{The same as in Fig. \ref{fig:59Fe_spectra_KB3G} for KB3G interaction and for the
$\beta$-decay transition $^{60}$Fe$(0^+)\to\,^{60}$Co($2^+$).}
\label{fig:60Fe_spectra_KB3G}
\end{figure*}

\begin{figure*}[!ht]
\centering
\includegraphics[width=\columnwidth]{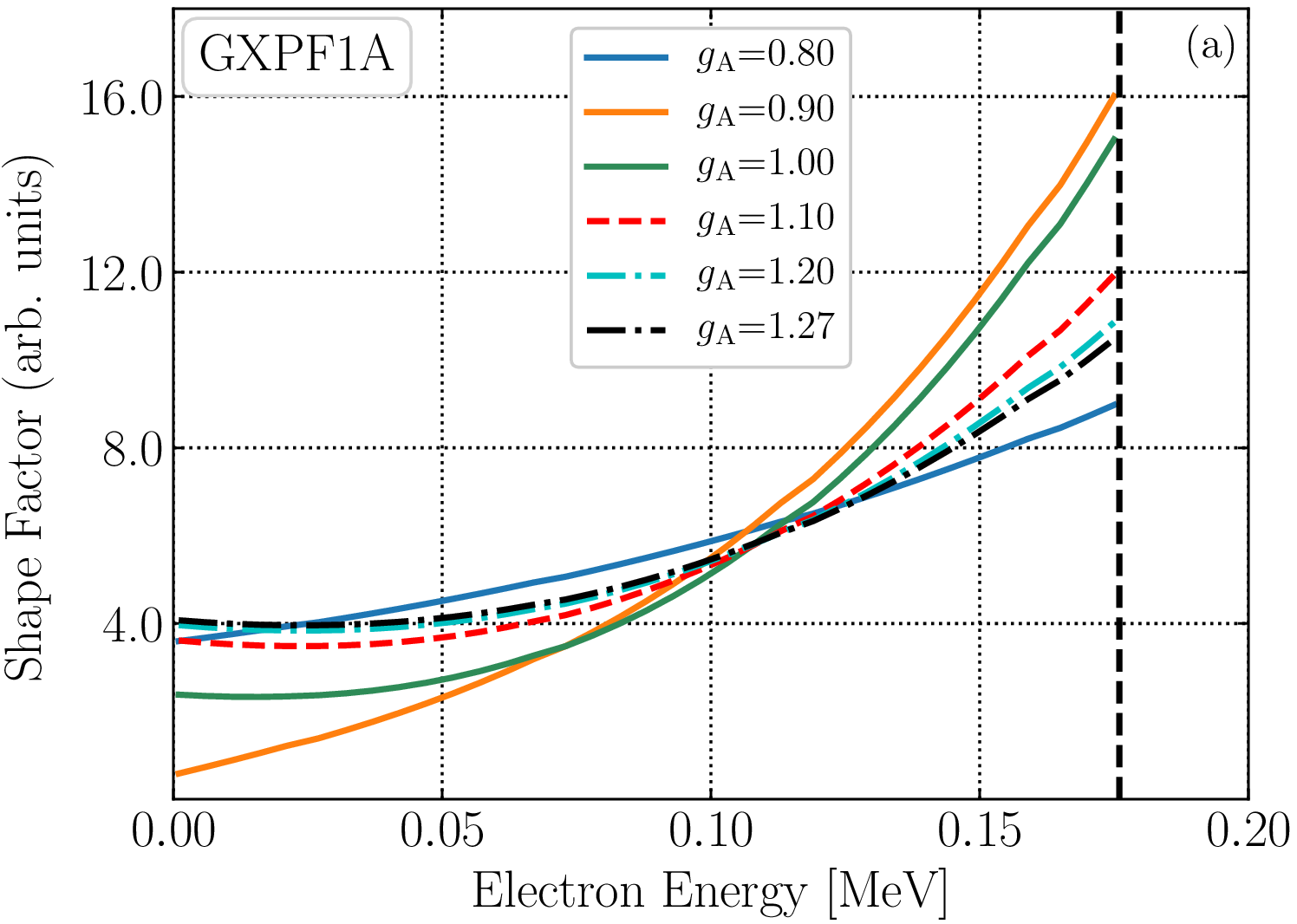}
\includegraphics[width=\columnwidth]{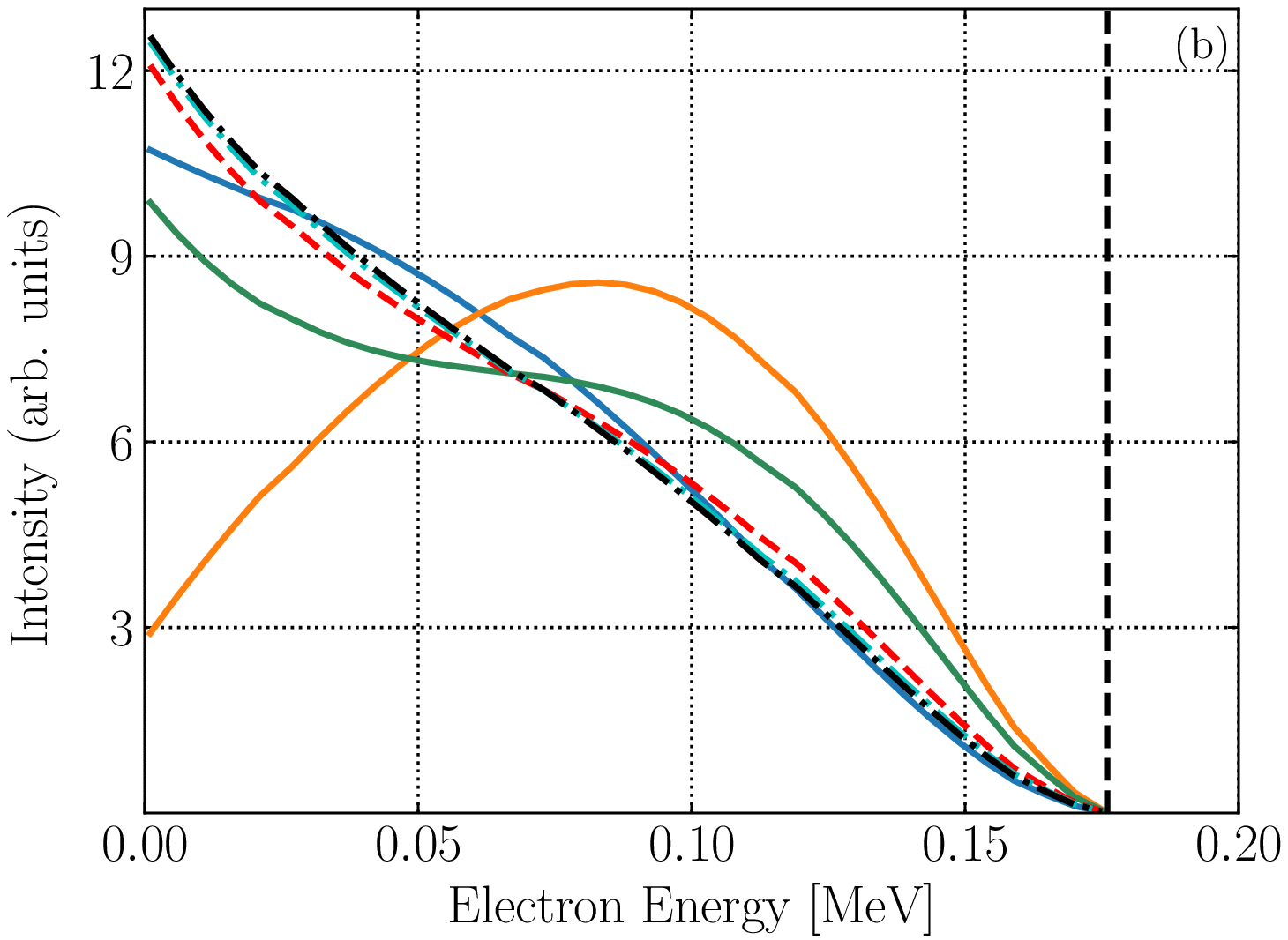}
\includegraphics[width=\columnwidth]{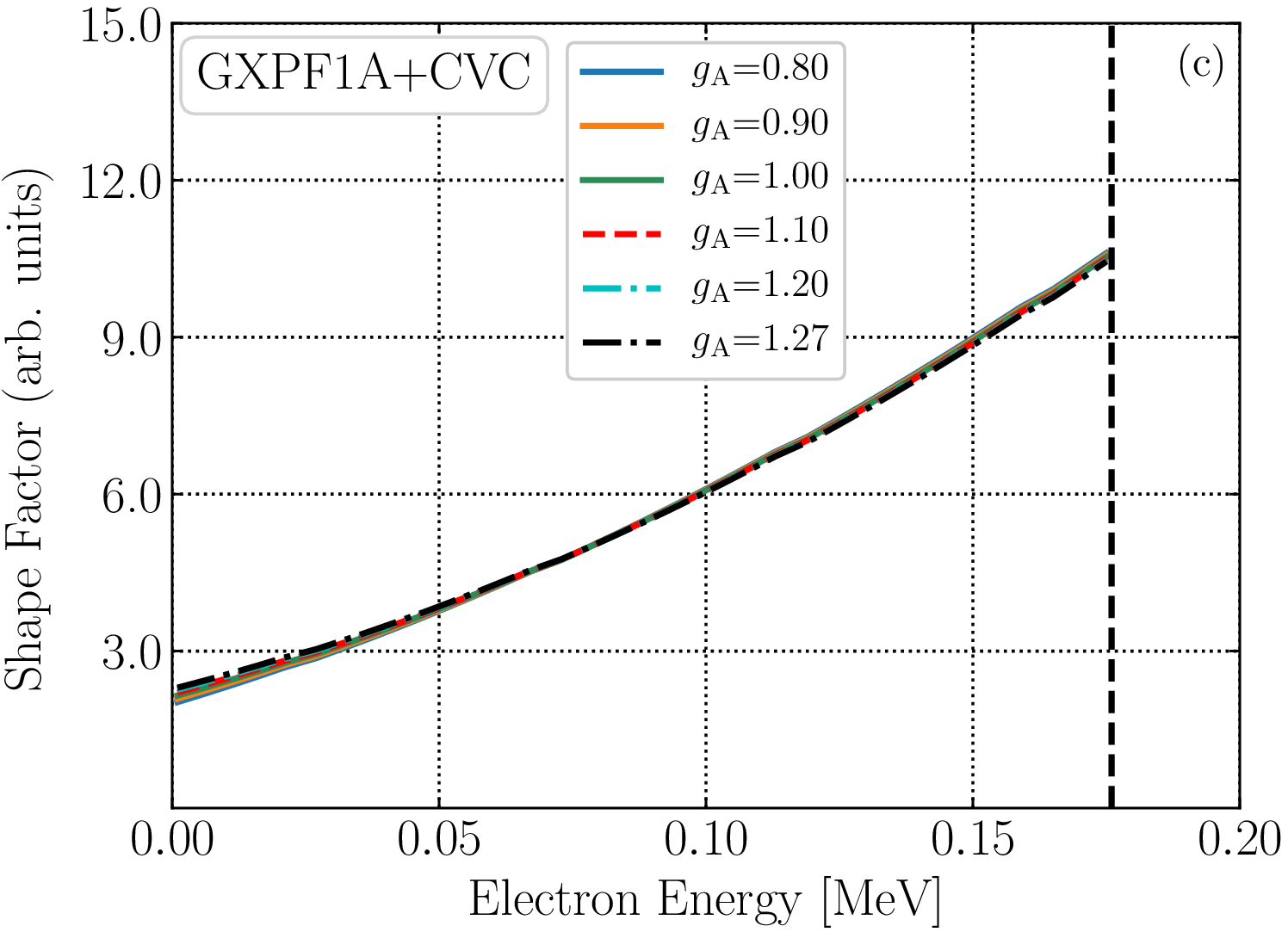}
\includegraphics[width=\columnwidth]{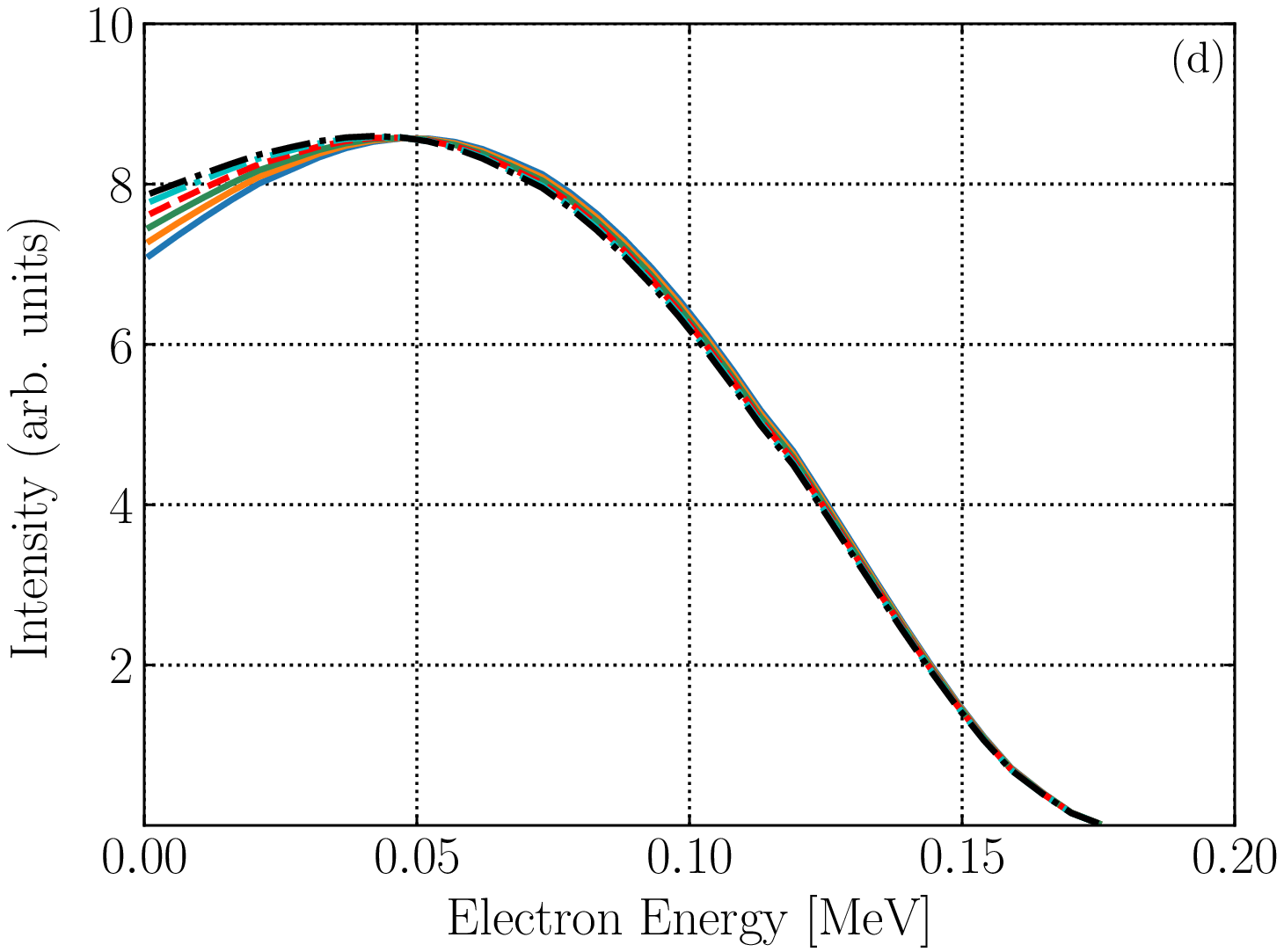}
\includegraphics[width=\columnwidth]{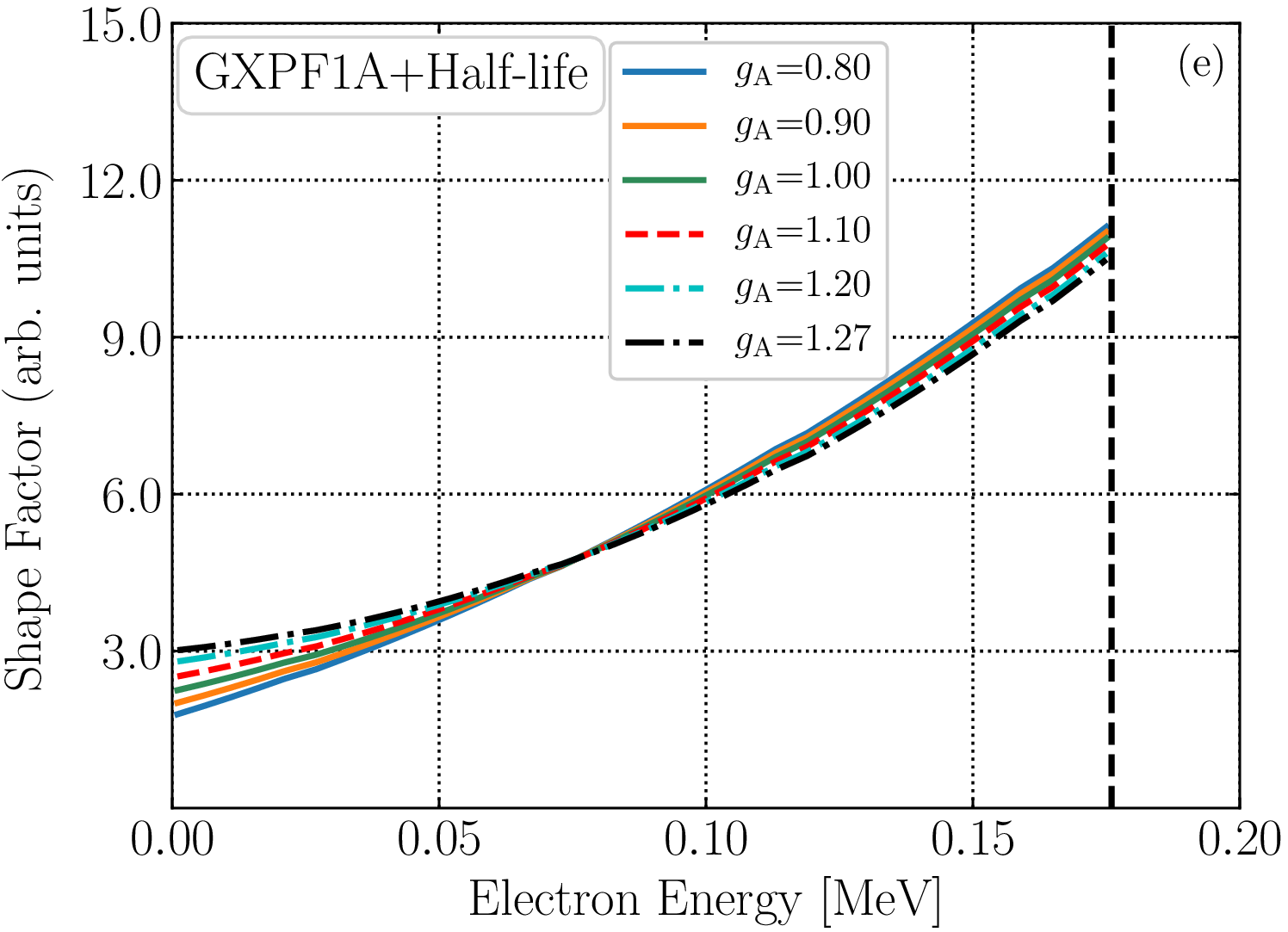}
\includegraphics[width=\columnwidth]{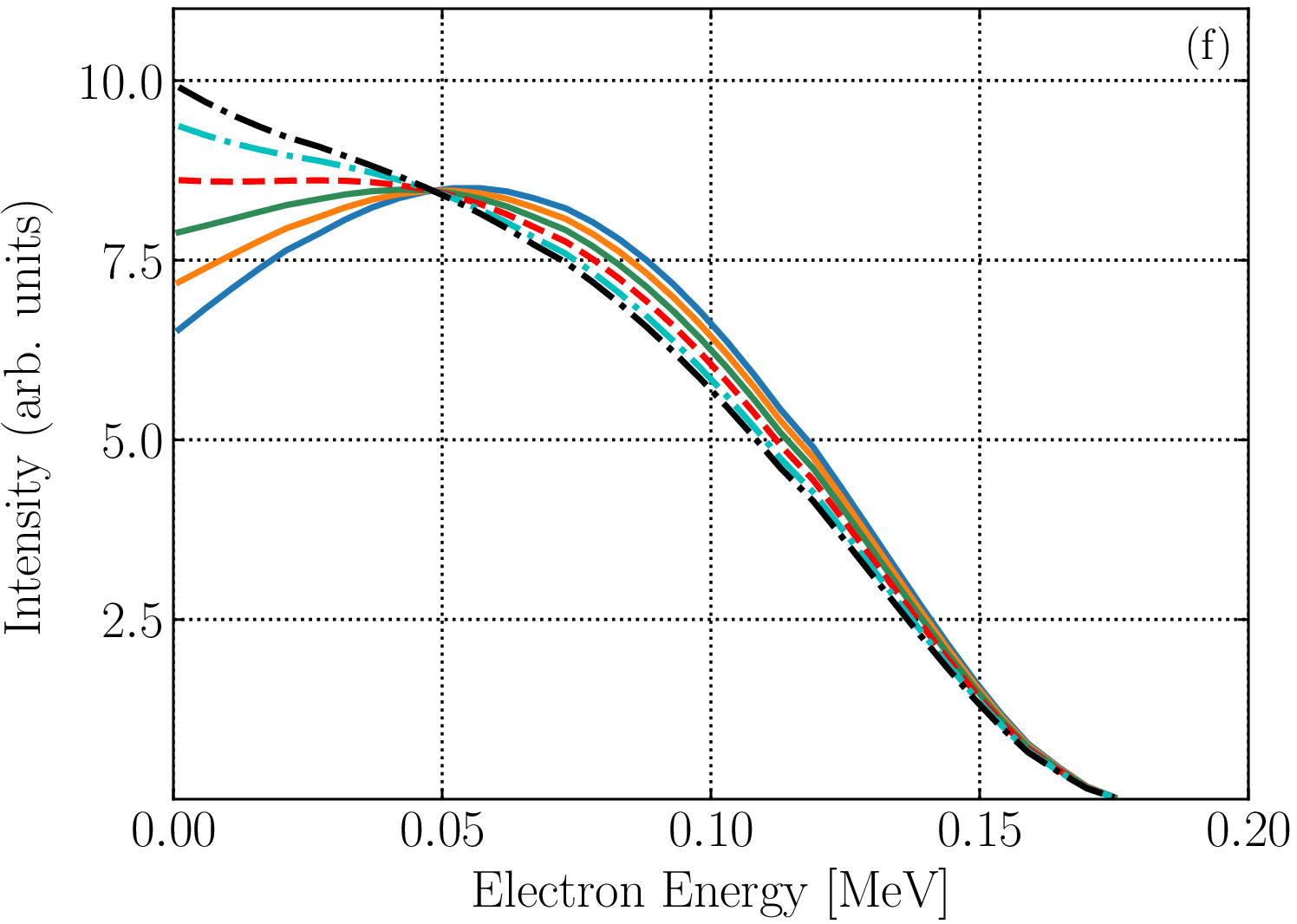}
\caption{The same as in Fig. \ref{fig:59Fe_spectra_KB3G} for GXPF1A interaction and for the
$\beta$-decay transition $^{60}$Fe$(0^+)\to\,^{60}$Co($2^+$).}\label{fig:60Fe_spectra_GXPF1A}
\end{figure*}

The computed shape factors (left panels) and electron spectra (right panels), corresponding to 
the decay transitions in $^{59,60}$Fe, are shown in 
Figs. \ref{fig:59Fe_spectra_KB3G}$-$\ref{fig:60Fe_spectra_GXPF1A} for the range 
$g_{\rm A}=0.80-1.27$ of the axial coupling. 
In these figures are presented the shape factor of Eq. (\ref{eq2}) and the electron spectra 
corresponding to the integrand of Eq. (\ref{tc}) as functions of the electron kinetic
energy, computed separately for the KB3G and GXPF1A interactions.
We have chosen to normalize the area under each curve to unity in order to facilitate 
easy comparison with the potential future data. 

The electron spectra of $^{59}$Fe are shown in the Figs. \ref{fig:59Fe_spectra_KB3G} 
and \ref{fig:59Fe_spectra_GXPF1A} for the KB3G and GXPF1A interactions, respectively. For
the bare interaction (our method 0)), the computed electron spectral shapes strongly 
depend on the value of $g_{\rm A}$. After constraining the value of the s-NME by the CVC 
relation (our enhanced method i)), the electron spectral shapes become almost insensitive 
to $g_{\rm A}$, while for the half-life-determined s-NMEs (our enhanced method ii)) the
spectrals shapes are moderately sensitive for both interactions. 
As seen in the figures, the spectra for each individual computation method show similar 
behaviour for both interactions. In particular for the method ii), of interest for the 
comparison with the potential future data, the electron spectral shapes computed with the two
interactions are consistent with each other. In this case, the differences of the spectra 
for different $g_{\rm A}$ values are more pronounced for the low-energy part of the spectra, 
below 0.5 MeV of electron energy, or for the middle part, between 0.7 MeV and 1.2 MeV of
electron energy. This low-energy part is the
harder part for the experiments to measure due to growing backgrounds which mask the
exact experimental shape. In the middle electron energy region above 0.7 MeV the differences
between the spectra are easier to resolve experimentally owing
to a smaller background contribution. It is our hope that the energy threshold of a
dedicated beta-decay experiment could go well below 0.5 MeV, as was the case for the
$^{113}$Cd experiments performed by Belli \textit{et al.} \cite{Belli2007} and by the
COBRA collaboration \cite{COBRA}. In the present case a potential additional difficulty
may be the small branching ratio, 0.18(4)\%, of the ground-state transition.

\begin{figure}[!ht]
\centering
\includegraphics[width=\columnwidth]{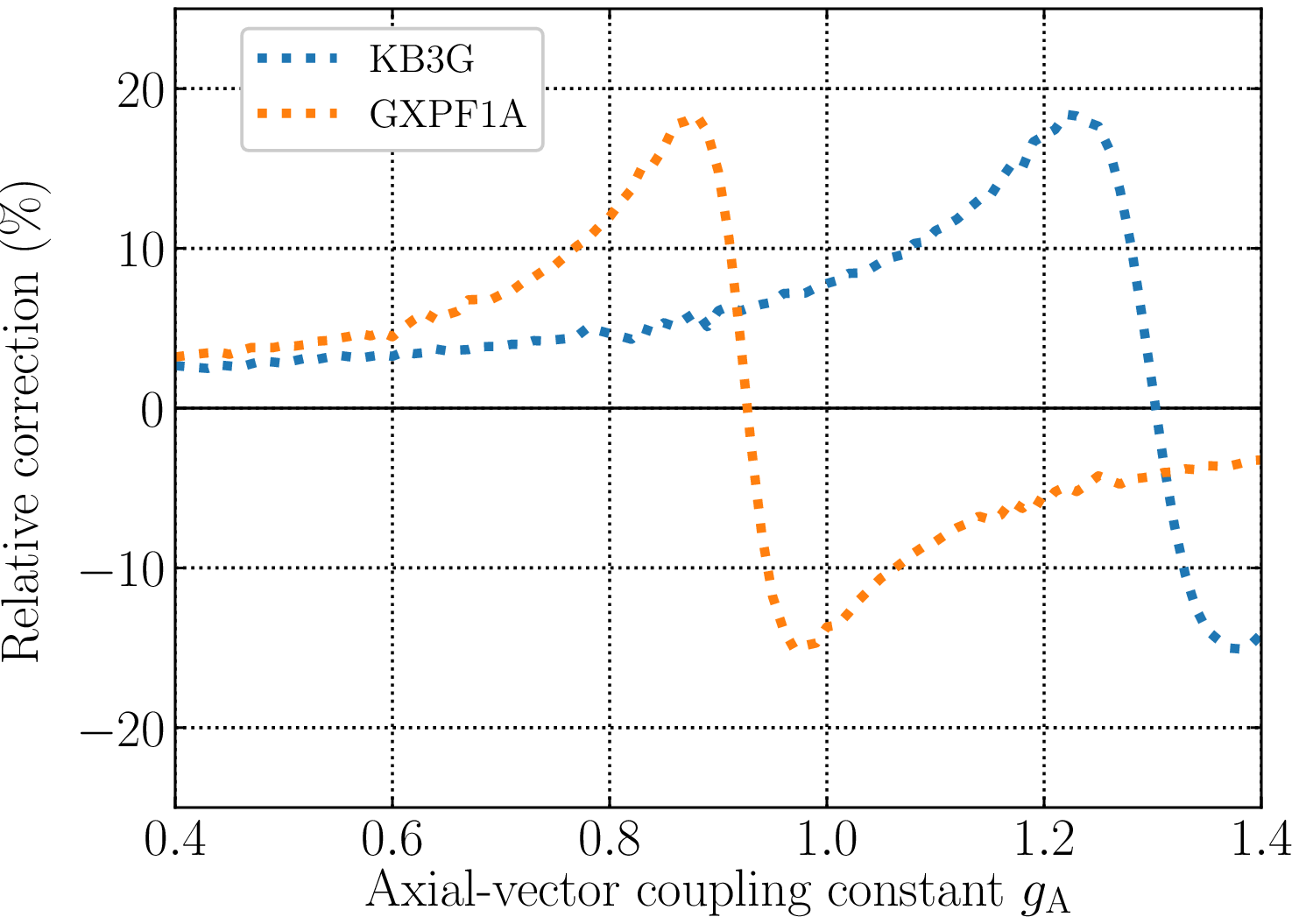}
\caption{Relative correction to the partial half-life for the $\beta^-$-decay of $^{60}$Fe 
stemming from the NLO corrections to the shape factor (i.e., with s-NME$=0$). The relative correction is presented
as a function of the value of the axial-vector coupling $g_{\rm A}$.}\label{fig:60Fe_HL_Corr}
\end{figure}

The electron spectra of the $\beta^-$-decay transition of $^{60}$Fe have previously been 
computed by Kostensalo \textit{et al}. in Ref. \cite{joel22017}. For this calculation, they 
have used the old Horie-Ogawa interaction, constrained to orbitals 
$\pi0f_{7/2}$, $\nu1p_{3/2}$, $\nu0f_{5/2}$ and $\nu1p_{1/2}$. They have performed the 
calculations for the electron spectra by using the pure shell-model-predicted matrix 
elements, without the CVC correction. From this interaction, the shape of electron spectra 
turns out to be independent of $g_{\rm A}$. In our case, we have used the complete 
$0\hbar\omega$ shell-model calculations in a full $fp$ 
model space with the well-established KB3G and GXPF1A interactions, and also constrained 
the value of the s-NME using the enhanced SSM procedures i) and ii). Our predicted shape 
factors and electron spectra for the $\beta$ transition of $^{60}$Fe, corresponding to the
KB3G and GXPF1A interactions, are shown in the Figs. \ref{fig:60Fe_spectra_KB3G} 
and \ref{fig:60Fe_spectra_GXPF1A}, respectively. The computed electron spectra from the bare 
interaction are found, again, to strongly depend on the value of $g_{\rm A}$. 
Interestingly enough, the predicted shape factors and electron spectra for the two
interactions are found to behave differently as functions of $g_{\rm A}$. This contradicting
behavior is resolved in the enhanced SSM procedures i) and ii). After fixing the value of
the s-NME, the $g_{\rm A}$ dependence of the electron spectral shapes is qualitatively 
similar for both used interactions. In the case of the method ii), relevant for the experimental
comparison, the spectral shapes are found to be moderately dependent on $g_{\rm A}$. The 
sensitivity to $g_{\rm A}$ is most pronounced for the low electron energies, below some 0.05 MeV.
Some $g_{\rm A}$ dependence is found also for the middle region of electron energies.

\begin{figure*}[!ht]
\centering
\includegraphics[width=\columnwidth]{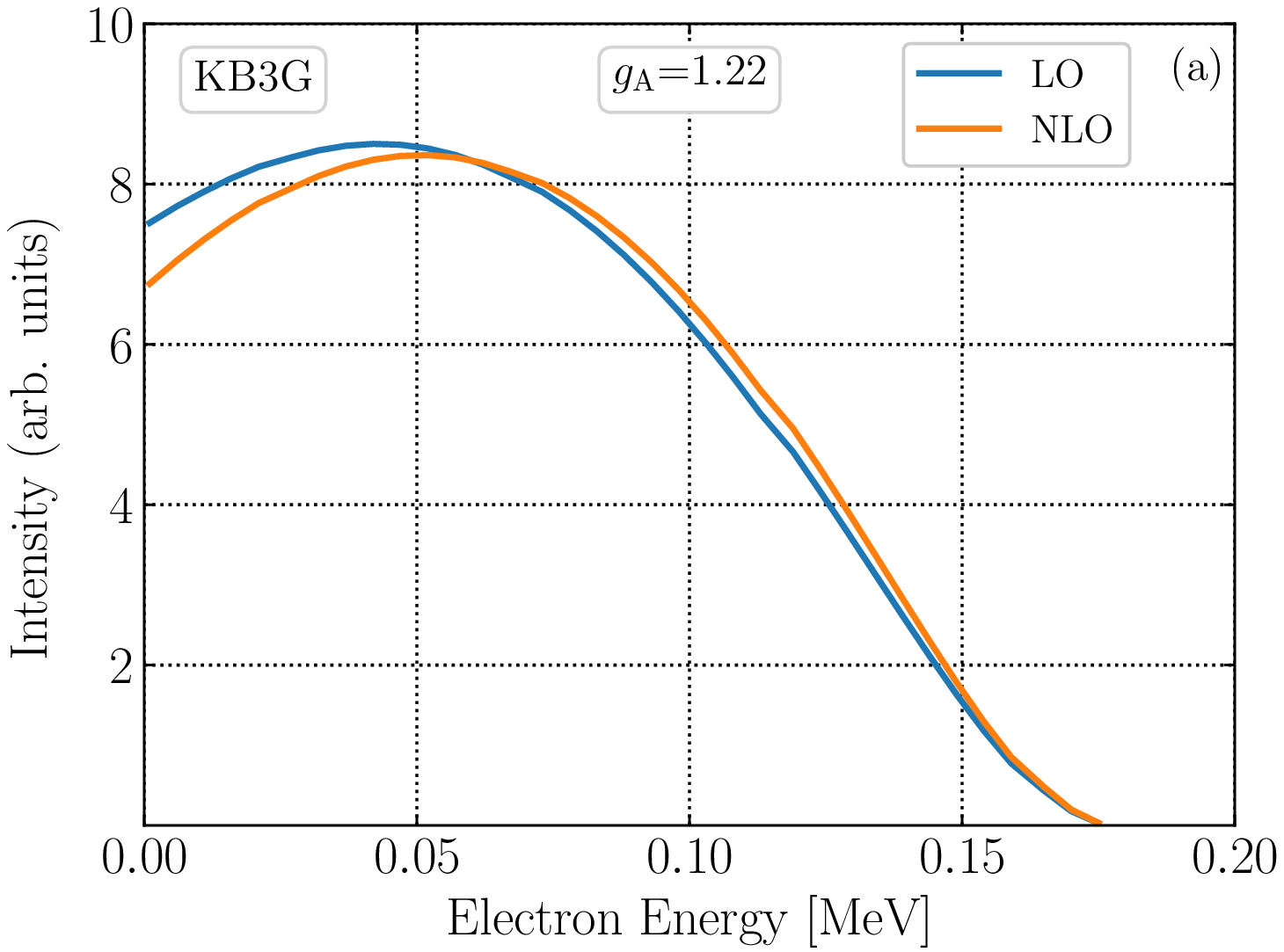}
\includegraphics[width=\columnwidth]{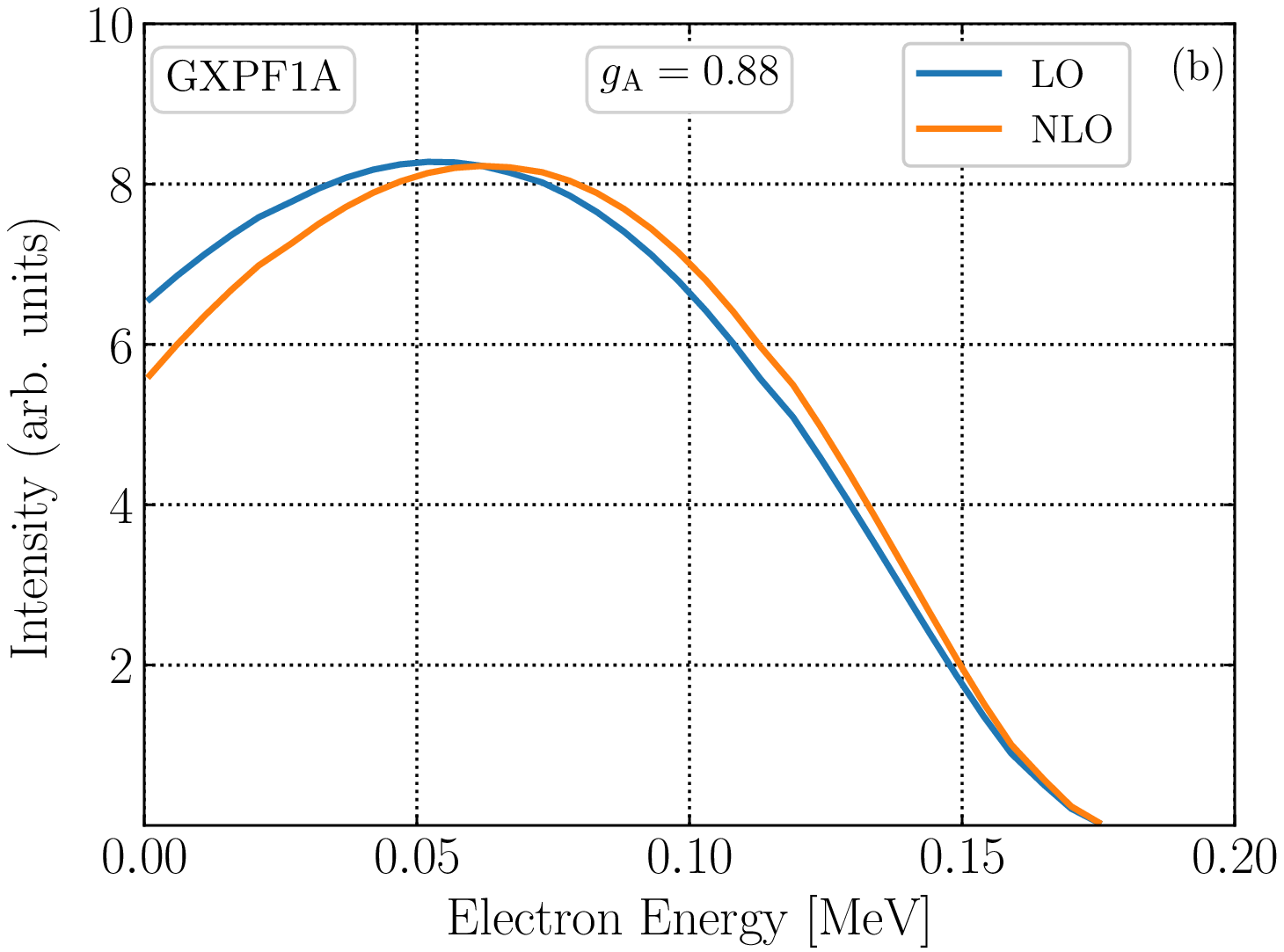}
\caption{Electron spectral shapes for the decay of $^{60}$Fe, computed using the bare 
KB3G (left panel) and GXPF1A (right panel) interactions (i.e., with s-NME$=0$). The 
indicated  $g_{\rm A}$ values of each panel correpond to those in Fig. \ref{fig:60Fe_HL_Corr}
for which the maximal NLO corrections to the partial half-life are achieved.}
\label{fig:60Fe_spectra_LO_NLO}
\end{figure*}

In Fig. \ref{fig:60Fe_HL_Corr}, in the case of the $^{60}$Fe decay, we present the relative 
NLO corrections to partial half-life as functions of the value of $g_{\rm A}$ for calculations
using the bare interactions, without any s-NME adjustments. As seen in this figure, the effects 
of the NLO corrections to the half-life are strongly dependent on the value of 
$g_{\rm A}$. Adding the NLO corrections leads at most to an about 18\% correction to the 
partial half-life, corresponding to the values of $g_{\rm A}=1.22$ and $g_{\rm A}=0.88$ for the 
KB3G and GXPF1A interactions, respectively. The partial half-life increases for $g_{\rm A}$
values below these values and decreases fast within a short interval after that. The NLO
corrections affect also the electron spectral shapes as seen in 
Fig. \ref{fig:60Fe_spectra_LO_NLO}, where the spectra are plotted at the above-mentioned
$g_{\rm A}$ values which give the maximum effects on the half-life in the bare-interaction
calculations. We can see from the figure that the effects of the NLO corrections are most 
significant at the low electron energies for both the KB3G and GXPF1A interactions.  
Interestingly enough, these cases are the extreme ones, and performing the 
s-NME adjustments according to the methods i) and ii) dampens the NLO effects considerably, 
and in these cases the partial half-lives and electron spectra collect negligible contributions 
from the NLO corrections to the $\beta$-decay shape function. 

\begin{table*}[!ht]
\caption{Decomposition of the dimensionless integrated shape function into vector 
$\tilde{C}_{\rm V}$, axial-vector $\tilde{C}_{\rm A}$, and mixed vector-axial-vector 
$\tilde{C}_{\rm VA}$ components for $g_{\rm V}=g_{\rm A}=1.00$.}\label{table:Fe_int_shape}
\begin{ruledtabular}
\begin{tabular}{lcccc}

   & & \multicolumn{2}{c} {$^{59}$Fe$(3/2^-)\to\,^{59}$Co($7/2^-$) } \B\\
\cline{2-5}
  & $\tilde{C}_{\rm V}$ & $\tilde{C}_{\rm A}$ & $\tilde{C}_{\rm VA}$  & $\tilde{C}$  \T\B\\
\hline

KB3G& 5.829$\times{10^{-6}}$ & 2.111$\times{10^{-6}}$ & -6.853$\times{10^{-6}}$ & 1.088$\times{10^{-6}}$\T \\
GXPF1A & 4.602$\times{10^{-6}}$ & 2.092$\times{10^{-6}}$ & -6.079$\times{10^{-6}}$ & 6.148$\times{10^{-7}}$ \\\\

KB3G+CVC& 1.410$\times{10^{-5}}$ & 2.111$\times{10^{-6}}$ & 9.795$\times{10^{-6}}$ & 2.600$\times{10^{-5}}$\T \\
GXPF1A+CVC & 1.110$\times{10^{-5}}$ & 2.092$\times{10^{-6}}$ & 8.677$\times{10^{-6}}$ & 2.187$\times{10^{-5}}$ \\\\

KB3G+Half-life& 4.623$\times{10^{-7}}$ & 2.111$\times{10^{-6}}$ & 3.716$\times{10^{-7}}$ & 2.945$\times{10^{-6}}$\T \\
GXPF1A+Half-life & 3.912$\times{10^{-7}}$ & 2.092$\times{10^{-6}}$ & 4.615$\times{10^{-7}}$ & 2.945$\times{10^{-6}}$ \\\\

   & & \multicolumn{2}{c} {$^{60}$Fe$(0^+)\to\,^{60}$Co($2^+)$ } \B\\
\cline{2-5}
  & $\tilde{C}_{\rm V}$ & $\tilde{C}_{\rm A}$ & $\tilde{C}_{\rm VA}$  & $\tilde{C}$  \T\B\\
\hline

KB3G& 1.199$\times{10^{-10}}$ & 7.123$\times{10^{-11}}$ & -1.845$\times{10^{-10}}$ & 6.561$\times{10^{-12}}$\T \\
GXPF1A & 1.368$\times{10^{-10}}$ & 1.606$\times{10^{-10}}$ & -2.961$\times{10^{-10}}$ & 1.381$\times{10^{-12}}$ \\\\

KB3G+CVC& 4.615$\times{10^{-10}}$ & 7.123$\times{10^{-11}}$ & 3.339$\times{10^{-10}}$ & 8.666$\times{10^{-10}}$\T \\
GXPF1A+CVC & 5.253$\times{10^{-10}}$ & 1.606$\times{10^{-10}}$ & 5.352$\times{10^{-10}}$ & 1.221$\times{10^{-9}}$ \\\\

KB3G+Half-life& 8.225$\times{10^{-12}}$ & 7.123$\times{10^{-11}}$ &-3.349$\times{10^{-12}}$ & 7.611$\times{10^{-11}}$\T \\
GXPF1A+Half-life & 2.144$\times{10^{-11}}$ & 1.606$\times{10^{-10}}$ & -1.060$\times{10^{-10}}$ & 7.611$\times{10^{-11}}$ \\\\

\end{tabular}
\end{ruledtabular}
\end{table*}

\subsection{Decomposition of the integrated shape function}

We have decomposed the integrated shape function to see the individual effects of the vector 
$\tilde{C}_{\rm V}$, axial-vector $\tilde{C}_{\rm A}$, and mixed vector-axial-vector 
$\tilde{C}_{\rm VA}$ components [see Eq. (\ref{intc})]. The integrated shape function $\tilde{C}$ 
and its decomposed components for the studied decay transitions, computed with the KB3G and 
GXPF1A interactions, are presented in Table \ref{table:Fe_int_shape} for $g_{\rm V}=g_{\rm A}=1.00$. 
For both interactions and studied transitions, the signs of the vector $\tilde{C}_{\rm V}$ and 
axial-vector $\tilde{C}_{\rm A}$ components are positive for all approaches of calculation. 

In the case of $^{59}$Fe, the mixed vector-axial-vector component is negative for the bare 
interaction calculations and positive for methods i) and ii). For both interactions, the vector 
contribution is larger than the axial-vector contribution in the calculations using the bare 
interaction and method i), while in the case of method ii) the axial part dominates. For the 
bare interaction calculation of $^{59}$Fe, the $\tilde{C}_{\rm A}$ component is about 36\% (KB3G) 
and 45\% (GXPF1A) of the vector component and the mixed component $\tilde{C}_{\rm VA}$ is slightly 
less than the sum of $\tilde{C}_{\rm V}$ and $\tilde{C}_{\rm A}$, and negative for both interactions. 
After constraining the s-NME, the decomposition for $^{59}$Fe is surprisingly similar for both 
interactions. In method i), for both interactions, the axial-vector component contributes little 
to the total integrated shape function, while in the method ii) the contribution of the vector 
component is very small for both interactions. 
In the case of $^{60}$Fe, for both interactions, the mixed component is negative for the  
bare-interaction and method ii) calculations, while positive for the method i). In the 
bare-interaction calculations, the vector and axial-vector components are roughly equally strong,
but for the method i) the vector part dominates and for the method ii) the axial part dominates.
One also notices that for the bare-interaction calculations the mixed component $\tilde{C}_{\rm VA}$ 
is negative and its magnitude is almost the sum of the $\tilde{C}_{\rm V}$ and $\tilde{C}_{\rm A}$ 
components. After constraining the value of s-NME, this degeneracy is lifted, in particular in the
case of the method ii).

\newpage

\begin{figure}[!ht]
\centering
\includegraphics[width=\columnwidth]{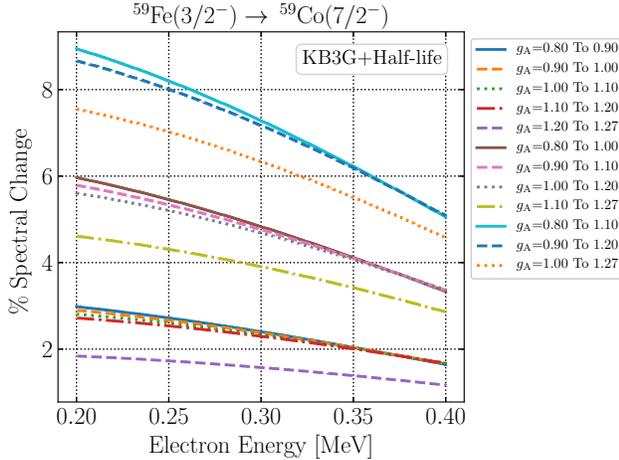}
\caption{ Percentual changes in the beta spectrum shape of Fig. 
\ref{fig:59Fe_spectra_KB3G}-(f), computed by using the interaction KB3G, in the energy region 
$0.2-0.4$ MeV. The spectral changes correspond to changes in the value of $g_{\rm A}$ by 0.1, 0.2, and 0.3 units within the interval $g_A = 0.8 - 1.27$.}\label{fig:per_KB3G.eps}
\end{figure}

 To give a rough idea of the needed experimental resolution in the determination of
the effective value of $g_{\rm A}$ from the beta spectra computed using the method ii), 
we have picked as an 
example the decay of $^{59}$Fe, computed with the KB3G interaction. The corresponding comparison is shown in Fig. \ref{fig:per_KB3G.eps}, where we have picked the energy region $0.2-0.4$ MeV of the beta spectrum of Fig. \ref{fig:59Fe_spectra_KB3G}-(f), since in this
interval the sensitivity of the spectrum to $g_{\rm A}$ is the strongest. In Fig. \ref{fig:per_KB3G.eps}  we show the change in the spectral shape in percents when changing $g_{\rm A}$ by 0.1, 0.2, and 0.3 units within the interval $g_A = 0.8 - 1.27$. One can see that when one wants one-unit resolution in
the value of $g_{\rm A}$ the corresponding change in he beta spectrum is of the order of $1-3$\%. For a two-unit or a three-unit resolution the corresponding spectral changes are in the ballpark of $4-6$\%~  and $6-8$\%, respectively. Similar percentual sensitivities of the beta spectra to the changes in the value of $g_{\rm A}$ are also recorded for the other cases treated by using method ii).

\subsection{Critical assessment of the present status of calculations} \label{critical}

We have presented in the previous sections the results of our search for
$g_{\rm A}$-sensitive electron spectral shapes in the $fp$ shell-model valence space. We have found two potential candidate transitions, $^{59}\textrm{Fe}(3/2^-)\to\,^{59}\textrm{Co}(7/2^-)$ and $^{59}\textrm{Fe}(3/2^-)\to\,^{59}\textrm{Co}(7/2^-)$, with some sensitivity in their spectral shapes to the variation in the value of $g_{\rm A}$. However, in order these cases to be useful for possible future measurements, it is may be necessary to improve the present calculations and access the systematic errors of them. The critical aspects of the present calculations relate to the determination of the values of the large vector NME $^V\mathcal{M}^{(0)}_{220}$ and the s-NME $^V\mathcal{M}^{(0)}_{211}$. First, the present calculations use only the $fp$ valence space leaving out possible particle-hole contributions from the adjacent shells, like the $gds$ shell above. These excitations could alter the value of the large NME, as also the values of the other NMEs, including the value of the s-NME, involved in the calculations. Second, the present way of determining the value of the s-NME through the CVC relation from the large NME [method i)] has uncontrolled uncertainties in it related to the fact that the present calculations are not "ideal", and subject to a severe truncation in the valence single-particle space. The method ii) uses the s-NME as a fit parameter, thus compensating the error in the present half-life calculation stemming from the limited valence space. In both methods, then many uncertainties can be present and extension of the presently used valence space is in order in possible future calculations. In this way one can at least have an estimate of the uncertainties related to the spectral-shape calculations.

A further complication in the present calculations are the different ways the many operators involved in the present calculations can be renormalized by the limited valence space and/or the true renormalization of the weak axial current. The renormalization can be different for the operators of axial and vector type, further complicating the interpretation of the present, small model-space, calculations. In this respect the forbidden unique transitions offer a much simpler approach since they are driven by a single axial NME and direct comparison of the computed and measured half-lives is in a position to make the determination of a "quenching factor" a rather straighforward procedure. Again, this quenching factor includes both the valence-space renormalization and the renormalization of the weak axial current. However, it could be imagined that in reasonably robust nuclear-model calculations in the future the two methods, SSM and the unique-forbidden scheme, could offer complementary information of the renormalization of the axial current and valence-space effects. These robust calculations should take into account, in addition to the presently included NLO corrections, also the effects of the two-body currents, as discussed recently in \cite{Gysbers2019}. The effect of these currents could be expectred to play a role of at least of similar magnitude as the NLO corrections to the nonunique forbidden beta decays.

\section{Conclusions}\label{Conclusions}

In this work we have performed a search in the $fp$ shell for possible forbidden nonunique beta transitions for which the associated beta spectrum shape depends notably on the value of the axial coupling $g_{\rm A}$, thus allowing the use of the spectrum-shape method in this context. We have found two candidate transitions, and we have performed a full $0\hbar\omega$ calculation of the shape factors and electron spectra for 
the second-forbidden nonunique $\beta^-$-decay transitions $^{59}$Fe$(3/2^-)\to\,^{59}$Co($7/2^-$) and 
$^{60}$Fe$(0^+)\to\,^{60}$Co($2^+$) in the full $fp$ single-particle space using the well-established effective shell-model interactions KB3G and GXPF1A. We include also the next-to-leading-order corrections to the $\beta$-decay shape factor. 

To test the predictive power of the adopted interactions, we have  calculated the low-lying energy spectra of the parent and daughter nuclei participating in the studied $\beta^-$ transitions and compared the results with the available data. The low-lying energy spectra agree reasonably with the  experimental data and only for $^{60}$Co we are unable to obtain the correct ground state. Also the spectroscopic properties are reasonably well predicted by these effective interactions. The obtained wave functions have been 
used for further calculations of the $\beta$-decay rates. 

In the present work, we have enhanced the original SSM by constraining the value of the small vector NME (s-NME) $^V\mathcal{M}^{(0)}_{KK-11}$ in two different ways: either by i) directly using a CVC relation, or ii) reproducing the experimental partial half-life by tuning the value of this matrix element. In the method ii), we have constraint the s-NME for each 
$g_{\rm A}$ separately. 
 The computed values of the s-NME are found to deviate by a factor of 2-3 between the two approaches, depending on the used interaction.
The evolution of the shape factors and electron spectra for the two second-forbidden nonunique $\beta^-$-decay transitions was studied within the interval of $g_{\rm A}=0.80-1.27$ using the two shell-model interactions. We found that the computed shape factors and electron spectra are sensitive to the value of the s-NME $^V\mathcal{M}^{(0)}_{KK-11}$. The enhancement method ii) is to be considered as the preferred one over i) since it can be used under the 
assumption of the impulse approximation and for restricted shell-model single-particle spaces. It also makes possible to reproduce the partial half-lives of the studied $\beta$-decay transitions in the calculations. At the end of the results section (Sec. III E) we have also taken a critical look at the deficiencies and the resulting inaccuracies of the present nuclear-model calculations.

In the case of the method ii), the shapes of the electron spectra depend 
somewhat on the value of $g_{\rm A}$ for both studied transitions, thus opening up
the possibility to use this revised SSM to access the effective value of 
$g_{\rm A}$ in these cases. 
For $^{60}$Fe the decay transition has a branching ratio of 100\%, making it a perfect candidate 
for future spectral-shape measurements and a subsequent application of the enhanced SSM. In order  
to see the origin of these variations in the electron spectral shapes we have decomposed the total integrated shape function $\tilde{C}$ into its vector 
$\tilde{C}_{\rm V}$, axial-vector $\tilde{C}_{\rm A}$, and mixed vector-axial-vector $\tilde{C}_{\rm VA}$ components.
The relative sizes and signs of these components lead to sensitivity or
non-sensitivity of the spectral shapes to the value of $g_{\rm A}$. We hope that in the future we can improve the characterization of the inaccuracies related to the used nuclear models in the context of the SSM. We nevertheless hope that our theoretical work strongly encourages future measurements of electron spectral shapes.

\section*{Acknowledgments}\label{Acknowledgments}

A. K. would like to thank  the Ministry of Human Resource Development (MHRD), Government of India, 
for the financial support for his Ph.D. thesis work.  P.C.S. acknowledges a research grant 
from SERB (India), CRG/2019/000556. J. S. has been partially supported  by the academy of 
Finland under the academy project no. 318043. 



\end{document}